\documentclass[emulateapj]{emulateapj}

\begin{document}
\shorttitle{Pulsar Disk Evolution}
\shortauthors{Currie \& Hansen}
\title{The Evolution of Protoplanetary Disks Around Millisecond Pulsars: The PSR 1257 +12 System}
\author{Thayne Currie\altaffilmark{1,2}}
\author{Brad Hansen\altaffilmark{2}}
\altaffiltext{1}{Harvard-Smithsonian Center for Astrophysics, 60 Garden St. Cambridge, MA 02140}
\altaffiltext{2}{Department of Physics and Astronomy, University of California-Los Angeles, Los Angeles, CA 90095} 
\email{tcurrie@cfa.harvard.edu}
\begin{abstract}
We model the evolution of protoplanetary disks surrounding millisecond pulsars, using PSR 1257+12 as a 
test case.  Initial conditions were 
chosen to correspond to initial angular momenta expected for supernova-fallback disks and disks formed 
from the tidal disruption of a companion star.  Models were run under two models for the viscous 
evolution of disks: fully viscous and layered accretion disk models.  Supernova-fallback disks 
result in a distribution of solids confined to within 1-2 AU and
produce the requisite material to form the three known planets surrounding PSR 1257+12. 
Tidal disruption disks tend to slightly underproduce solids interior to 1 AU, required for forming 
the pulsar planets, while overproducing the amount of solids where
no body, lunar mass or greater, exists.
Disks evolving under 'layered' accretion spread somewhat less and deposit a higher 
column density of solids into the disk.
In all cases, circumpulsar gas dissipates on $\lesssim 10^{5}$ year timescales, 
making formation of gas giant planets highly unlikely. 
\end{abstract}

\section{Introduction}
The planetary system orbiting the pulsar B1257+12 first reported
in 1992, (Wolszczan \& Frail, 1992) and subsequently confirmed (Wolszczan, 1994
 \& 2000), was the first extrasolar planetary system discovered. By the time of the 
 discovery of the first planet around a solar-type star (Mayor \& Queloz 1995; Marcy \& Butler 1996)
 the planetary nature of the system had already been confirmed and elucidated by the detection
of the signal of resonant perturbations between the two largest planets in the system (Wolszczan 1994). 
The eccentricities of all three known planets are now known to be extremely low ($e < 0.026$), and
the orbits of the outer two planets are nearly coplanar (relative inclination
  $< 6^{o}$) (Konacki \& Wolszczan, 2003, hereafter KW03).  Table 1 lists the orbital parameters 
the three planets (from Table 2 of KW03).
Wolszczan (1996) has also argued for the presence of a fourth signature from
the pulsar timing data, though its mass and semimajor axis are uncertain.

However, despite this conspicuous head start, there is still very little attention devoted to the theory of
planet formation in systems of this type. Such neglect is no doubt due to the fact that the initial
 conditions for planet formation around
a pulsar must necessarily be very different than in the case usually considered in planetary formation
theories. In particular, the protoplanetary disk in this case represents an adjunct to stellar death, rather than
stellar birth. It is important to note, however, that the very different provenances need not, in fact,
give rise to very different conditions. Most theories for how a millisecond pulsar might acquire a
protoplanetary disk (reviewed in Podsiadlowski 1993) result in a gaseous disk in orbit about a central gravitating
body of roughly $1 M_{\odot}$. Most such theories even describe a disk of approximately solar composition
(although some, such as that of Livio, Pringle \& Saffer 1992, produce a disk very poor in hydrogen
and helium). 

This situation is sufficiently similar to the standard picture that it may be analysed with
the same tools as are normally used in the study of planet formation.
Of course, there are still some fundamental differences 
between the situation we discuss here and the traditional one.
Most importantly, the planets are formed from an expanding disk since the original extent of the
disk is $\lesssim$ several AU (the justification for this is described in Section 2). Only that
 material which expands out (as the result of viscous evolution) to the distances of the current
planets is important for planet formation. This is different from the normal case, where much of the
mass is added to the disk at large radii as the result of accretion from a protostellar cloud.  
Nevertheless, the same fundamental physics behind viscous protoplanetary disk evolution should still
apply.

In this paper we investigate the evolution of the gaseous protoplanetary disk around a pulsar, incorporating
a detailed description of the physical state of the gas and how this affects the disk evolution.
This can be regarded as an extension of Phinney \& Hansen (1993) and Hansen (2000), using more
detailed models and incorporating modifications to the standard model for the evolution of gaseous proto
planetary disks, specifically the "layered disk" model (Gammie 1996; Armitage, Livio \& Pringle
2001). 

The structure of our paper is as follows.  In Section 2 we describe the 
two likely scenarios for producing protoplanetary disks surrounding millisecond pulsars, the 
supernova fallback and tidal-disruption scenarios, and discuss the initial disk conditions 
that result from them.  In Section 3 we discuss the 
standard model for viscous
protoplanetary disk evolution as well as modifications to it, incorporating the layered disk 
accretion theory.  We describe the numerical procedure and initial conditions for 
our model runs in Section 4.  Section 5 details the gas and dust column density ($\Sigma_{g}$ \& $\Sigma_{d}$)
and disk central temperature ($T_{c}$) evolution assuming a supernova-fallback disk, comparing
results for a fully viscous disk model with a layered disk model.  Section 6 presents results for 
the disk evolution assuming a tidal-disruption disk.  In Section 7, we summarize our results 
describe future work to model the formation of planets around millisecond pulsars.

\section{Sources for Protoplanetary Disks around Millisecond Pulsars}
Two leading formation scenarios for protoplanetary disks around pulsars are now investigated: the 
supernova-fallback and tidal disruption scenarios.  These models 
 span a wide range of initial disk angular momenta: $\sim 10^{49}$-$10^{52}$ ergs $\cdot$s.
\subsection{Supernova-Fallback Disks}
The simplest model for the formation of a protoplanetary disk around a pulsar is the supernova-fallback scenario 
whereby some fraction of the outer shell mass of the supernova fails to escape the gravitational potential of the 
surviving neutron star.  This material then settles into a disk surrounding the pulsar.  Chevalier (1989), 
Hashimoto et al. (1989) and Lin et al. (1991) modeled 
supernova fallback, arguing that $\lesssim 0.1$ $M_{\odot}$ of material may remain.  
The initial angular momentum for such a disk is rather low, typically 
$\sim 10^{49}$ ergs $\cdot$ s (Menou et al. 2001).  

Similarity solutions used to 
describe the structure of a fallback disk were derived by Menou et al. (2001) 
and were used to argue for confinement of the disk 
to sub-AU scales ($\sim 10^{12}$ cm).   However, 
these solutions assume a balance between heating and cooling in determining the 
temperature structure of fallback disks, an assumption generally more appropriate for disks 
with far larger initial angular momenta.  We take a different approach, 
allowing for imbalance of heating and cooling as well as 
advection in determining the disk temperature evolution.
The fallback material is assumed to be metal-rich, and the total mass is set at $\sim$ 
$10^{-3}$ $M_{\odot}$, consistent with the fallback mass of Type II supernovae (Chevalier 1989).

\subsection{Tidal-disruption Disks}
Alternatively, the disk may form from interactions with a binary companion.  
We base our assumption for the initial distribution of disk material on the initial angular momentum
expected from scenario 5 in Phinney \& Hansen (1993): the circumpulsar disk material originates
from a companion star that was tidally disrupted after the supernova of the primary.  
This would occur if the off-center
supernova recoil either kicked it into or close to a binary companion or the post-SN orbital
parameters for the secondary resulted in tidal disruption. 

There are some indications that tidal disruption can be a possible explanation for the origin of a 
circumpulsar disk.  
An analysis of post-SN orbital parameters from Kalogera (1996) demonstrate that the survival rate of 
post-SN binaries is much greater for small initial orbital separations.  Post-SN parameters favor 
a binary separation equal to or lesser than the initial separation and also favor more highly eccentric orbits, factors 
which make tidal disruption possible.  If tidal disruption occurred as the 
pulsar kicked into its companion at a high velocity then such 
a pulsar would likely have a high proper motion.  The PSR 1257+12 system exhibits an extremely 
high space velocity of 
$\sim$ 300 km $s^{-1}$, consistent with this disk formation scenario.  
The resulting circumpulsar disk would have an initial angular momentum 
$J \sim 10^{51}$--$10^{52} ergs\cdot s$.   The disk composition is assumed to be solar. 

\section{Viscous Disc Surface Density and Temperature Evolution}
Our models begin with an initial protoplanetary disk confined to small radii ($\leq 3 AU$),
which should be true of both scenarios for disk formation (e.g. Phinney \& Hansen 1993).
Viscous evolution in this disk results in the accretion of most of the material, but some of the
mass must flow outwards to conserve angular momentum (Lynden-Bell \& Pringle 1974). Eventually some
of this material condenses out into solids and may thus provide the seed material for the 
observed planets. The goal of our calculations is to examine whether there is sufficient material
in the disk to plausibly explain the observed planetary mass. An additional constraint is whether
the conditions in the disk are appropriate for the incorporation of dust and solids into planetesimals
that would decouple from the gas and begin the process of planet accumulation. This
is primarily a temperature threshold -- if the gas temperature is too high most heavy elements
will remain in a gaseous form.

 Phinney \& Hansen (1993) examined these issues
 with simple $\alpha$-disk models for a range of initial angular momenta. The results
were inconclusive because of two competing effects. For high angular momentum budgets there was
indeed enough mass on large scales, but it remained too hot to plausibly deposit solid material
until the disk had expanded to scales of several AU, somewhat larger than the observed planetary
system. On the other hand, lower angular momentum budgets resulted in the gas becoming cool and
depositing mass on
smaller scales -- too small to really explain the observed planets. Hansen (2000) observed that
a potential solution to this problem was to modify the disk evolution to take account of a
viscosity that decreases at low temperature -- a hypothesis consistent with many treatments of
traditional protoplanetary nebulae (Gammie 1996; Sano \& Miyama 1999). We now investigate this
idea in more detail.
\subsection{Standard Model}
Since the initial disk extent was probably small ($\lesssim$ several AU) we model the viscous 
disk evolution using a full diffusion prescription instead of calculating similarity solutions 
as in Menou et al. (2001).  
Under the thin disk approximation, the viscous evolution of the surface density $\Sigma$ is given by 
the following equation (e.g. Pringle 1981; Hartmann 
1998):
\begin {equation}
\frac{\partial \Sigma}{\partial t} 
= \frac{3}{r} \frac{\partial}{\partial r}[r^{1/2}
\frac{\partial}{\partial r}(\nu \Sigma r^{1/2})], \label{Vis1}
\end {equation}
where $\nu$ is the viscosity. This viscosity is determined by the 
microphysics of the gas and is affected by the midplane temperature.
The viscosity is parameterized as an $\alpha$ viscosity (Shakura and Sunyaev,1973)
\begin{equation}
\nu=\alpha c_{s}^{2}/\Omega
\end{equation}
where $\alpha$, a constant which relates the efficiency of viscous transport, is set to
 $10^{-2}$ for MHD-driven turbulent flow. 

Viscous heating, irradiation, radiative cooling, and radial advection of material through the 
disk determine of the disk temperature evolution 
\begin {equation}
\frac{\partial T_{c}}{\partial t} 
= \frac{2}{\Sigma c_{p}}(Q_{+} +Q_{irr} - Q_{-})
 - v_{r}\frac{\partial T_{c}}{\partial r} \label{Temp1}
\end {equation}
where $c_{p}$ is the disc specific heat ($c_{p}\sim 2.7 \frac{\Re}{\mu}$), $\Re$ is the gas constant, 
and $\mu\sim 2.3$ is the mean molecular weight for temperatures $\sim 10^{3}$K.  $Q_{+}$ is the viscous heating rate given as
\begin {equation}
Q_{+} = \frac{9}{8}\nu \Sigma \Omega^{2}, \label{Qp1}
\end {equation}
and $Q_{irr}$ is the heating rate from irradiation (due to accretion onto the pulsar) 
\begin{equation}
Q_{irr}=\frac{L_{acc}}{4\pi r^{2}}(1-\beta)cos \phi, \label{Qirr1}
\end{equation}
where cos$\phi$=$\frac{dH}{dr}-\frac{H}{r}$ and the albedo ($\beta$) is 0.5
(Frank, King, and Raine 2002; see also King and Ritter 1998).
We assume that $L_{acc}$ is the 
minimum of the accretion luminosity onto the pulsar and the Eddington luminosity, and 
allow for self-shadowing of outer regions of the disk by puffed-up inner regions.
$Q_{-}$ is the radiative cooling term given as 
\begin {equation}
Q_{-} = \sigma T_{e}^{4},
\end {equation}
and the vertically averaged radial velocity of material through a cross section
in the disc is 
\begin {equation}
v_{r}=-\frac{3}{\Sigma r^{1/2}}\frac{\partial}{\partial r}(\nu \Sigma r^{1/2}). \label{Vr1}
\end {equation}

The cooling rate can be cast in terms of the midplane temperature $T_{c}$ by relating 
$T_{c}$ to the local effective temperature ($T_{e}$) through the condition of vertical radiative transport. 
 From Hubeny (1990) we find that in the optically thick limit
\begin{equation}
T^{4}=\frac{3}{8}\tau T_{e}^{4}
\end{equation}
where $\tau = \frac{\Sigma}{2}\kappa$ and for the optically thin limit
\begin{equation}
T^{4}\sim T_{e}^{4}/\tau.
\end{equation}

To complete the microphysical description of the gas, we compute the opacity $\kappa$ as a function
of temperature and density. We use the frequency-averaged opacities from Bell \& Lin (1994),
shown in Table 2 with slight modifications. The 
analytical fits to these opacities have the form of $\kappa = \kappa_{o}\rho^{a}T^{b}$.  The temperature
range over which a given analytical fit of $\kappa$ is valid is determined by requiring continuity of
the function $\kappa$ between different domains.
 For low temperatures, the temperature range over
which a given power-law prescription is valid is independent of density.   When the disc is partially ionized
there is a non-negligible dependence on the density and thus $\Sigma$ for the boundaries: 
\begin{equation}
T_{max,n}
=[\frac{\kappa_{n}}{\kappa_{n+1}}(\Sigma\Omega(\frac{\Re}{\mu})^{1/2})^{a_{n}-a_{n-1}}]^{\frac{1}{b_{n+1}-b_{n}+1/2(a_{n}-a_{n+1})}}
\end{equation}
where the midplane gas density is given as $\rho_{c}=\Sigma c_{s}/\Omega$ and the adiabatic sound 
speed is given as $c_{s}=\sqrt{\Re T_{c}/\mu}$.  The opacities in Table 2 assume solar abundance.  
Menou et al. (2001) note that opacities for species expected in supernova-fallback disks are similar 
to those for a solar mixture.  

The analytical fits from Bell \& Lin (1994) compare favorably with more sophisticated numerical 
calculations (e.g. Semenov et al. 2003; Ferguson et al. 2005) over most temperature 
ranges.  However, they underpredict the opacity by as much as three orders of magnitude in the n=4-5 regimes 
as can be seen from Semonov et al. (2003; at log(T)$\sim$ 3.25). By the n=6 ($\kappa$$\propto$$T^{10}$, $\sim$ 
log(T)=3.5) regime, the opacities are 
again consistent with other calculations.  Semenov et al. cite an improper truncation of the molecular 
water opacity as the likely source for this discrepancy.  Because numerical calculations from several different groups 
(e.g. Pollack et al. 1994; Alexander \& Ferguson 1994; 
Semonov et al. 2003) all find a much higher opacity, there is strong evidence that the Bell \& Lin opacities are too low.  
Therefore, when a disk region is in the n=4-5 opacity regimes we set a 'floor' to the opacity equal to opacity at 
transition temperature between n=4 and 5, where the Bell \& Lin predictions are consistent with others.  Thus, over 
a small temperature range the opacity is roughly constant and comparable to the values of 
Semenov et al. and others.  

\subsection{Modifications for Layered Disk Evolution}

Extensive work over the last decade has shown the effectiveness of the magnetorotational instability (MRI)
in providing the anomalous viscosity for accretion disks in a variety of astrophysical 
situations (Balbus \& Hawley 1991). However, one situation where its usefulness is debatable is
precisely the one of interest in this case -- a protoplanetary disk. 
This may be especially true in the SN fallback case where the cooling disk may quickly become neutral and thus 
unable to drive MHD turbulence (Lin et al. 1991; Menou et al. 2001). 
The principal issue is the 
ionization state of the gas which is an exponential function of the gas temperature (see Gammie 1996). When the disk gas is only partially ionized, the coupling between magnetic field 
and disk gas is weakened and eventually the MRI ceases to operate. The gas temperatures in the outer parts
of protoplanetary disks are perhaps a few hundred Kelvin, so that ionization is expected to be very
low and the viscosity may be considerably reduced.

This requires that we modify the disk evolution described in Section 3.1.  In particular, we adopt
the formalism of Gammie (1996), in which accretion is assumed to occur only through a thin `active' layer on
the surface of the disk.  Ionization is maintained by irradiation, either by cosmic rays or X-rays
from the central star, and so the MRI can still operate.  The rest of the material is 
assigned to an inviscid, unevolving `dead' layer.
Thus, at each radius we must consider both an active surface density $\Sigma_a$ as well as a total
surface density $\Sigma$. The relative extent of the active layer is a function of the magnetic
Reynolds number in the midplane (Fleming \& Stone 2003), with the transition occurring over a finite
temperature range.  

The ionization fraction for solar abundances is dominated by collisionally-ionized potassium (Umebayashi 1983) and varies 
over five orders of magnitude from 1000K to 800K.  Thus, for tidal disruption disks we assume that the disk is fully viscous 
at 1000K and transitions (linear in log T space) to a layered state by 800K.  The treatment for the supernova-fallback disk 
is more difficult since such disks are unlikely to have a solar composition of elements and also may be initially hot enough 
to trigger nucleosynthesis.  To address this issue, we follow Fujimoto et al. (2001) who used the pre-SN chemical 
composition models of a 20 $M_{\odot}$ progenitor from Nomoto \& Hashimoto (1988) to model the initial conditions and early 
evolution of fallback disks.   Nucleosynthesis in the disk is possible at stellocentric distances 
$\le$ 3$\times10^{3}$ $r_{g}$, where r$_{g}$ is the Schwarzschild radius 
($\frac{2GM}{c^{2}}$), which corresponds to $\sim 2\times10^{10}$ cm for a 20$M_{\odot}$ progenitor.  This distance is 
probably too small to be relevant for planet formation from fallback disks.  Regions beyond this 
should have a similar composition to that of the ejected layers of the pre-SN star.  The exact composition of these layers 
varies strongly.  Fujimoto et al. considers three layers: a silicon rich, oxygen rich, and helium-rich layer.  We assume that 
these layers are well mixed in the disk and calculate the mean composition of the disk assuming that all layers 
contribute equally.  The dominant gas-phase elements are O, Si, He, and S;  $\sim 26\%$ are metals.  Because a sizeable fraction 
of metals are retained (including, presumably, potassium) we assume that the disk transitions from fully viscous to layered 
over the same 1000 to 800K range.

In both the fallback and tidal-disruption disk models, we assume that $\Sigma_{a}$ has a minimum of 200 g$cm^{-2}$ 
which is twice the stopping surface density for cosmic rays using interstellar medium values (100 g$cm^{-2}$ for 
two disk faces).  The stopping depth for x-rays is a much smaller $\sim 0.1 g cm^{-2}$.

Thus, for layered accretion equations 1, 3, and 6 are now modified such that
\begin {equation}
\frac{\partial \Sigma}{\partial t} 
= \frac{3}{r} \frac{\partial}{\partial r}[r^{1/2}
\frac{\partial}{\partial r}(\nu \Sigma_{a} r^{1/2})],
\end {equation}
\begin {equation}
Q_{+} = \frac{9}{8}\nu \Sigma_{a} \Omega^{2}
\end {equation}
and
\begin {equation}
v_{r}=-\frac{3}{\Sigma r^{1/2}}\frac{\partial}{\partial r}(\nu \Sigma_{a} r^{1/2}). 
\end {equation}

The vertically averaged radial velocity of the disc, $v_{r}$, becomes smaller as the disc evolves into
a layered state since the energy transported by the disk active layers is shared by
the entire disk bringing the dead zone into thermal equilibrium with the base of the active layer.
The temperature change due to advection is then reduced by $\Sigma_{a}/\Sigma$ in the
layered disk region.  
The optical depth, $\tau$, is now given as $\tau = \frac{\Sigma_{a}}{2}\kappa$.

The small $\Sigma_{a}$ 
value may underestimate the true column density of material transported.  The cosmic-ray flux at the center 
of a supernova remnant may in fact be significantly larger than typical interstellar medium values.  
Furthermore, while most of the disk may be magnetically dead, it is not clear that the same column of material is 
mechanically dead: turbulent eddies in the active region may overshoot the active/dead boundary and transport 
some mass from the dead zone (see Fleming \& Stone 2003).  Therefore, the fully viscous and layered 
prescriptions adopted here should be seen limiting cases for viscous transport in circumpulsar disks, 
where the viscous evolution actually manifest in circumpulsar disks is likely somewhere in between. 

\subsection{Solid Body Evolution}
Modeling only the gas evolution is not sufficient for investigating possible planet formation out of such
a disk as a fraction of the original disk mass will condense into solids once the disk cools past 
certain temperature thresholds.  Here we describe a crude model for tracking the solid body evolution
of an originally all-gaseous circumpulsar disk.  We assume that once the disk at some stellocentric
distance drops below a temperature threshold it removes a fraction of its mass to solids in the next
timestep.  This neglects dust dynamics in the evolution of the circumpulsar disk.
Specifically, we first have neglected the time required for grains to settle to the disk midplane.  As this depends
critically on the grain size as well as the level of anisotropy of disk turbulence - the component of turbulence
acting vertically must be extremely low for complete settling (e.g. Dubrulle et al. 1995)- we make the simplifying
assumption that the solid bodies can settle to the disk midplane on short timescales and that turbulence is unable
to stir particles to large scale heights.  
Highly anisotropic 
hydrodynamical forms of turbulence (e.g. Youdin \& Chiang 2004) or layered disk models operating under MRI turbulence
 (e.g. Currie 2005) should allow grains to settle to a thin scale height.  Second, we neglect the effect of radial
migration.  Such migration may alter the dust-to-gas ratio in disks, trigger planetesimal formation 
(Youdin \& Shu 2002), and perhaps even alter the efficiency of planet formation (Currie 2005).  However, as we shall 
see, gas from circumpulsar disks typically depletes on very short, $\sim 10^{5}$ year timescales.  Furthermore, as the disk 
cools very large columns of dust, much greater than the Minimum Mass Solar Nebula (MMSN; Hayashi 1981) may be deposited for some models.  Both of these conditions hasten planetesimal formation 
by gravitational instability such that grain radial migration may not be a strong effect.
Therefore, while simple, this model should still allow one to get a zeroth-order understanding 
of the expected regions of planet formation from a circumpulsar disk.

The first temperature threshold we use occurs at $\sim 500 K$ which takes into account the condensation of metal
grains which should finish around $\sim 470 K$.  Once a gridpoint drops below $500 K$ we remove a fraction of its
mass from the gas disk and place it into a solid body distribution:
\begin{equation}
\Sigma_{d,metals}=c\times \Sigma_{g}.
\end{equation}
Because the supernova fallback disk is comprised of 26\% metals, c=0.26 in the supernova fallback model. 
For the solar-composition tidal disruption model, we set c=0.01.

The second temperature threshold, relevant for the tidal-disruption model, occurs at the 
water ice condensation point.  This transition occurs roughly at $170 K$ 
and removes from the disk a larger fraction of mass than the condensation of metals.  Specifically, when a 
gridpoint drops below $170 K$ we then remove the following mass from the gas disk and deposit it into the solid
body distribution
\begin{equation}
\Sigma_{d,ice}=0.032\times \Sigma_{g}.
\end{equation}
hese temperatures are significantly less than $\sim 1000$ K, below which the disk 
is likely magnetically dead and unable to drive turbulence in the disk midplane.  

Mass can also be added to the solid body distribution by advection of material from a hot region of the disk across
a temperature threshold to a cooler region.  We crudely model this by ascribing a certain fraction (0.01 or 0.26 for across
the metal grain threshold and 0.032 for the water ice threshold) of disk material designated for solids once said
material crosses a threshold.  We then set this fraction to be zero for all regions for which the disk is cool 
enough to allow condensation.  If, after a timestep, that grid point to the cooler side of the threshold now has
some 'solids' we add that amount to the solid body distribution for that grid point and subtract it
from the calculated gas distribution.

We also track the amount of solid material interior to $1 AU$ and exterior to 1 (2) $AU$ for the 
supernova fallback model (tidal disruption model).  As the three known planets
are all within $1 AU$ of the pulsar we want to see how much mass is contained in this region as a 
function of time.  Specifically, we want to see if, in order to form the known planets, we must 'import' material
from beyond $1 AU$ or require that they formed much further out and migrated inward.
Models for the fourth signature from this system suggest the existence of a Uranian to Jovian mass body with a large
semimajor axis ($\sim 5 AU$ or larger).  We test how likely such formation would be in a circumpulsar disk in our
formation scenario by tracking the mass of the disk exterior to $ 2 AU$.  

\section{Numerical Procedure}

We solve for the evolution of the surface density and temperature profiles
using an explicit finite difference method over 250 equally spaced radial grid points.  
The timestep taken is small at the beginning because of limits of the Courant
condition for numerical stability, but it is eventually expanded such that we cover a time evolution
of the disc over $10^{6}$ yrs.  To guard against numerical
 instabilities arising at the boundaries between opacity regions we smooth the 
values for the opacity at such boundaries to be a composite of the opacity on
 either side of the boundary.

We follow the evolution of circumpulsar disks after the point at which it has cooled to 
$\lesssim$ 10,000 K, prior to which it expands rapidly.
Formation scenarios for circumpulsar disks typically start with very compact disks 
($\lesssim$ 0.1 AU) and $\sim$ 10$^{49-52}$ ergs$\cdot$s.  
 Early tests found that starting the tidal-disruption disk runs with material spread over 
$\sim$ 0.1 AU scales caused the disk to heat up to $\gtrsim$ 10$^{6}$ K until it spread to 
$\sim$ several AU in less than 1 year, where it cooled to $\lesssim$ 10,000 K.  The lower 
mass of the supernova fallback disk results in smaller, more transient spreading such that  
a fallback disk cools to $\lesssim$ 10,000 K after spreading for $<$ 1 AU.
Thus, we uniformly spread the disk material
for our initial surface density profile over 0.3 AU (3 AU) for the supernova-fallback  
(tidal-disruption) models.  Initial temperature for the disc is
set at $\sim$ 3500 K though we find that for our initial surface density
the temperature quickly reaches values independent of the 
starting condition.  

We present results for the supernova-fallback scenario and then the tidal-disruption scenario.  
For each scenario we do model runs for full viscous disk models and then layered disk models.  
For the latter models, we use the active/
dead layer surface density ratios from equations 10 and 11.  
We start with initial angular momentum values of $10^{49}$, $10^{51}$, and $10^{52} ergs\cdot s$, corresponding 
to the supernova fallback model and two tidal-disruption models, respectively.
The goal is to run the disc evolution for reasonable initial models to 
see what constraints can be put on the planet formation from such disks.  
We compare our results with the masses and positions of the three known 
planets in PSR 1257+12, and place constraints on the locations and positions 
of any other planetary mass bodies in the system.  Important parameters for 
these constraints are the evolution of $\Sigma_{g}$, the gas surface density, 
$T_{c}$, the central temperature, and $\Sigma_{d}$, the solid mass 
surface density.  We also compute the absolute value of the mass accretion rate throughout the disk ($\dot M$) 
and track the Toomre Q parameter ($Q_{grav}$) to see the disk may periodically become gravitatonally 
unstable.
\section{Supernova Fallback Models}
\subsection{Fully-Viscous Accretion}
Figure 1 shows the gas surface density ($\Sigma_{g}$) and central temperature ($T_{c}$) profile 
evolution for the fully-viscous model, assuming that the disk originates as supernova-fallback material. 
The disk evolves on a brisk $10^{4}$ year timescale.  By $\sim 10^{5}$ years, the surface density of gas
drops to below 1/100th that expected for the Minimum Mass Solar Nebula.  For comparison, circumstellar 
dust and gaseous disk components are expected to evolve on longer, $\sim 10^{6}-10^{7}$ yr timescales (e.g. 
Currie et al. 2007a; Zuckerman et al. 1995), after which only gas-poor/free debris disks remain (e.g. 
Hernandez et al. 2006; Currie et al. 2007b).
Circumpulsar disks evolve on shorter timescales largely 
because the initial angular momentum of a supernova-fallback disk is $\gtrsim$ 100-1000 times smaller.  
A high disk temperature, and thus high rate of viscous transport, results from the compact fallback disk.  
Furthermore, irradiation from accretion onto the neutron star heats up the inner disk regions to very high 
($\gtrsim$ 10$^{5}$ K) temperatures during the early evolutionary stages (t$\lesssim$ 1000 yr), causing 
the disk to spread rapidly.  Irradiation impacts the disk at very acute angles and the disk cools after 
it spreads.  This allows all but the innermost disk regions to become self shadowed quite easily after 
$\sim$ 10$^{3}$ yr.  Viscous heating is then required to maintain high disk temperatures.
This results in a fast cooling of the disk from $10^{3}$-$10^{5}$ years. 

Despite the fast gas accretion timescale, the column density of solids, $\Sigma_{d}$, is still large enough 
to form approximately Earth mass bodies if the final assembly of planets is very efficient, owing to
the rapid cooling of disk material to below 500K (Figure 2) and advection across the 500 K threshold.  Typical 
$\Sigma_{d}$ values are $\sim$ 10 MMSN values, up to $\sim$ 100$\times$ MMSN at $\sim$ 0.1-0.5 AU, 
even though by $\sim 10^{5}$ years the \textit{gas} 
column density is depleted by a factor of $\gtrsim$ 1000.  

In this model, formation of Earth-sized bodies is highly unlikely outside 1 AU, 
though it may be possible form asteroid-sized bodies at this distance.
There are some suggestions that the presence of such a body is consistent with the 
pulsar frequency derivatives obtained by KW03 as suggested by Konacki \& Wolszczan, in prep.   
About 8.6 $M_{\oplus}$ of solids exist interior to 1 AU by $\sim$$10^{4}$ years.
Thus, forming the known planets of PSR 1257+12 from this reservoir of solids requires that the 
accretion process be $\gtrsim$ 95\% efficient.  Since the gas dissipation timescale is fast, 
the formation of gas giant planets from a supernova fallback disk is unlikely.

Thus, the likely result from this formation scenario and disk model is a system of one or more Earth-mass 
 bodies, orbiting very close to the star with no planets or circumstellar debris beyond $\sim$ 1 AU.  In the 
next section we see if this result for the supernova-fallback model is sensitive to our viscosity prescription: 
in other words, does the resulting distribution of solid mass differ if the disk spends much of its time in a 
layered state?
\subsection{Layered Accretion}
When the disk is allowed to evolve in a layered state the resulting evolution timescale is 
slightly longer than the fully viscous case, $\sim$ 
$10^{5}$ years (Figure 3).  The disk temperature characteristically remains higher than in the 
fully viscous model through $\sim$ 10$^{4}$ years.  
The $\Sigma_{d}$ profile follows a very different pattern, showing pronounced peaks at $\sim$ 0.5 AU 
and then later at 0.2 AU (Figure 4).  
The simulation was run several times with slightly varying initial parameters, and the same peak structure 
was reproduced each time.  This peak is due to a pileup of material at the dead zone/outer active zone boundary. 
If the disk material cools below 500K before gravitational instability 
is triggered (e.g. Armitage, Livio, \& Pringle 2001), then a substantial mass of solids can condense out of the 
disk.  The very small disk angular momentum, high $\Sigma_{d}$, and assumption of layered accretion combine to 
produce these peaks in the gas, and in turn dust, column densities.   
The mass of solids deposited within 1 AU of the pulsar can be enormous under the layered accretion assumption. 
 Figure 5 suggests that 30 $M_{\oplus}$ of solids can condense out of the disk interior to 1 AU, more 
than enough to form the known planets orbiting PSR 1257+12.  Interestingly, in both the fully viscous and 
layered disk models, most of the solid material sediments out at roughly the same place, $\sim$ 0.2-0.5 AU, 
which is conspicuously similar to the current locations of the known pulsar planets.  
Because a layered disk spreads so 
slowly, far less solid material is deposited at distances $\ge$ 1 AU (Figure 5, righthand side).  Under this 
model, any large bodies forming at $\sim$ 1 AU must be $\lesssim$ 0.1-0.5 $M_{\oplus}$ in mass.  In both the 
inner ($\le$ 1 AU) and outer ($\ge$ 1 AU) disk regions, the formation of gas giant planets by core 
accretion is highly unlikely given the very fast gas dissipation timescale (e.g. Figures 3 and 5).  

Thus, a supernova-fallback disk evolving by layered accretion will likely produce several high-mass (
$\sim$ 1-10 $M_{\oplus}$) bodies 
within $\sim$ 1 AU and may form bodies the mass of asteroids or Kuiper belt objects beyond $\sim$ 1 AU.  
This resulting architecture is consistent with the known planets in the PSR 1257+12 system 
but inconsistent with the existence any Uranian or Jovian-mass bodies orbiting at large semimajor axes.
In the next section we investigate disk evolution with initial conditions set by the tidal 
disruption models to see if a different disk formation mechanism is also consistent with the 
PSR 1257+12 system.
\section{Tidal Disruption Models}
\subsection{Fully-Viscous Accretion}
The first notable difference between the supernova fallback and tidal-disruption models is that the latter tend to have 
much larger disks, owing to their larger initial mass and angular momentum.  Figures 6-8 and 9-11 respectively 
show the results for a 
fully viscous disk with J$\sim 10^{51}$ and $J\sim 10^{52}$ ergs$\cdot$s of initial angular momentum.  
In the first case, by $\sim 10^{3}-10^{4}$ years, 
the disk expands to $\sim$ 8 AU.  Gas depletes on $\sim$ $10^{4}$ year timescales (Figure 6), and  
solid body densities comparable to MMSN values emerge by $\sim$ $10^{4}$ years with most of the mass within $\sim$ 3 AU (Figure 7).  
Initially, the disk has a very high accretion rate of 
$\sim 10^{-6}$ $M_{\odot}$ $yr^{-1}$, similar to that for classical T Tauri stars, but falls by more than two orders of magnitude 
within $\sim 10^{4}$ years.   Just over 8 $M_{\oplus}$ of material, slightly less than needed to form the pulsar planets,
 resides interior to 1 AU by $10^{5}$ years; a large amount $\sim 150 M_{\oplus}$ is deposited outside of 2 AU.  

A disk with $J\sim 10^{52}$ ergs$\cdot$s of initial angular momentum evolves on similar timescales (Figure 9).  
The $\Sigma_{d}$ profile shows that slightly less (more) mass is deposited interior (exterior) to 1 (2) AU than 
in the $J\sim 10^{51}$ ergs$\cdot$s case.  This is due to a slightly longer viscous evolution.  The solid mass 
exterior to 2 AU (Figure 11) exhibits a 'sawtooth' like pattern through $10^{4}$ years, 
owing to evaporation of solids as the outer disk heats up and then cools from expansion.  The direct heating of the disk from 
irradiation also contributed to evaporation of solids, especially in the first $\sim$ $10^{4}$ years.
The in-situ formation of $\gtrsim$ 5 $M_{\oplus}$-mass planets is still difficult interior to 1 AU.
About 300 M$_{\oplus}$ of solids exist beyond 2 AU by $\sim$ 10$^{5}$ years.
The formation of gas giants is, again, highly unlikely because circumpulsar 
gas depletes to below 10 $M_{\oplus}$ by 2$\times 10^{5}$ years.

In general, we find that the amount of solid mass interior to 1 AU is just under 8 $M_{\oplus}$ and the 
amount exterior to 2 AU is $\sim$ 100-300 $M_{\oplus}$.  It is then plausible that many 1-10 $M_{\oplus}$ bodies
 may be able to form beyond $\sim$ 1-2 AU in the tidal-disruption scenario 
for circumpulsar disk formation, though this formation.  However, since the disk falls below 
$\sim$ 1000K everywhere except for the innermost regions by $\sim 10^{3}-10^{4}$ years, the resulting solid body 
profiles (Figure 7,10), and expected locations for planet formation (Figure 8,11), 
may strongly depend on the fully viscous assumption.  
We now investigate what happens if we allow the disk to evolve by layered accretion.
\subsection{Layered Accretion}
Differences in how $\Sigma_{g}$ and other parameters evolve in the layered (compared to fully viscous) case are more 
pronounced for tidal-disruption disk models since the higher initial angular momentum and mass of the disk results 
in a larger difference between the actively accreting column of gas $\Sigma_{a}$ and the total column.  This is 
evident in the $J\sim10^{51}$ ergs$\cdot$s case as shown in Figure 12.  By $\sim$ 500 years, the disk between 
0.5 and 3 AU has evolved into a layered state.  $\Sigma_{g}$ in the 'layered' region is relatively higher than 
in the fully viscous case, owing to the pile up of gas onto the layered region/outer fully viscous region boundary 
that is subsequently redistributed throughout the disk interior to the boundary.  Over the next $\sim 10^{5}$ years 
the size of the layered region shrinks as material piles up onto its outer edge, eventually triggering gravitational 
instability of the disk at this region.  The instability, however, is marginal and operates only long enough to 
redistribute some mass quickly: planet formation by gravitational instability (Boss 2005) is highly unlikely.  Both 
$\Sigma_{g}$ and $T_{c}$ evolve on $\sim 10^{5}$ year timescales.  The long $T_{c}$ timescale results from 
a small advection term (from $\Sigma_{a} << \Sigma_{g}$)  and a near equilibrium that is reached by the heating ($Q_{+}$) 
and cooling ($Q_{-}$) terms in the $T_{c}$ evolution equation.  It is not until the advection term becomes important 
again that the imbalance between heating and cooling becomes upset and the central temperature declines substantially.

The mass accretion rate through the disk follows a constant to inverted structure characteristic of layered accretion (Figure 13).  
While the accretion rate is several orders of magnitude less than in the corresponding fully viscous case it stays above 
$\dot M\sim10^{-10} M_{\odot} yr^{-1}$ for $\ge 10^{4}$ years, or somewhat longer than in the fully viscous model.  
The resulting $\Sigma_{d}$ profile drops from $\sim$ 500 to 1000 yr, owing to viscous heating and irradiation.  
Eventually the disk cools and the final $\Sigma_{d}$ value settles to a 
median density of $\sim$ 30-40 g cm$^{-2}$ between 1 and 4 AU by $\sim$ 10$^{5}$ years (Figure 13).  
The amount of solid material interior to 
1 AU and exterior to 2 AU is set by $\sim 10^{5}$ years (Figure 14).  The amount of solid material 
deposited within 1 AU is $\sim$ 7 $M_{\oplus}$, or not quite enough to account for the three known planets; 
the bulk of the mass is deposited beyond $\sim$ 2 AU.
Even though up to 200 $M_{\oplus}$ of solids exist beyond 2 AU it is highly unlikely that gas giant planets can form around 
millisecond pulsars since the gas dissipates on $\sim 10^{5}$ year timescales.  

Layered disk models initially with $J\sim 10^{52}$ ergs$\cdot$s, though with a 
larger initial mass than $J\sim 10^{51}$ ergs$\cdot$s models, evolve on similar 
($\sim 10^{5}$ year) timescales (Figure 15).  The disk stays in a layered state from $\sim$ 10$^{3}$ to 10$^{5}$ years.  
$\Sigma_{d}$ evolves in a similar 
way to the other layered disk run though the solid column density above $\sim$ 10 gcm$^{-2}$ extends to 6 AU instead of 4 AU as with the 
previous run.  The densities are
$\sim$ 10$\times$ MMSN column densities through 6 AU (Figure 16).  This is because the disk, initially $\sim 0.1 M_{\odot}$ is 
confined to a smaller initial radial spread than for the solar nebula.  
The mass of solids deposited both interior to 1 AU and exterior to 2 AU is higher than in the fully viscous case by about 
a factor of two (Figure 17).  In general, it then appears that layered accretion disks 
deposit a higher mass of solids into the disk than fully viscous disks.
The circumpulsar gas drops below $\sim$ 10$M_{\oplus}$ by $\sim$ 10$^{5}$ years, so that no model predicts the presence of 
circumpulsar gas at ages comparable to those observed around pre-main sequence stars ($\sim$ 10$^{6}$-10$^{7}$ years).
Thus, the $\lesssim$$10^{5}$ year gas dissipation timescale in all model runs imposes
draconian requirements on the efficiency of gas giant planet formation even for models that swiftly form 
the cores of gas giant planets (Rafikov 2004; Currie 2005; Alibert et al. 2005). Such planet formation is 
effectively ruled out for both fallback and tidal disruption circumpulsar disk formation scenarios.
Although true gas giants are unlikely to form, there is sufficient
mass to allow for the formation of massive rocky or ice giant planets
with no gas envelopes.  The persistence of circumpulsar gas may also
still strongly affect the dynamics of planetesimals in the feeding zones of growing 
planets (Rafikov 2004; Currie, Kenyon, \& Bromley in prep.) and thus affect the final outcome of the pulsar planet formation 
process.  

We can now summarize our expectations for planet formation around millisecond pulsars when the tidal-disruption model is assumed
 for the circumpulsar disk origin.  In most models not quite enough material within 1 AU forms to account for all the 
planetary mass in the system; planet formation is preferred beyond $\sim$ 1 AU scales.
A Uranus-mass body orbiting at large stellocentric distances is consistent with results from pulsar timing 
methods, but the formation of a body with a much more massive gaseous envelope (e.g. a Jovian planet) from either the supernova-fallback 
or tidal-disruption scenarios is extremely unlikely.  
\section{Summary \& Future Work}
We have modeled the evolution of protoplanetary disks surrounding millisecond pulsars.  We investigated two formation 
scenarios for these disks, supernova fallback and tidal disruption, under two models for the disk viscous evolution: 
a fully viscous model and layered accretion model.  

For the supernova fallback scenario, the gas density drops well below MMSN values (at $\sim$ 1 AU) 
by $\sim 10^{4}$ - $10^{5}$ years while by $\sim 10^{5}$ years the solid mass 
is $\sim 10-100$ times that of the MMSN values within $\sim$ 0.5 AU of the pulsar.  
The most likely outcome for supernova fallback disks 
is a system of Earth mass bodies confined to within 1 AU of the pulsar, a system architecture consistent with the 
PSR 1257+12 system.  For the tidal disruption scenario, gas dissipates on $\lesssim$ 10$^{5}$ year timescales, and
the solid body distribution reaches 1-10$\times$ MMSN values by $10^{4}$ years 
with little material extending beyond $\approx$ 6 AU.  Most of the mass is deposited beyond $\sim$ 1-2 AU.
The most likely outcome from this scenario is a system of Earth-mass bodies between 2 and 6 AU but little material beyond 
this distance.  Not quite enough solid mass exists to form the known planets in the PSR 1257+12 system in situ.
While both models can account for the formation of Earth-mass planets somewhere in the disk, the supernova fallback model favors 
formation of such planets interior to $\sim$ 1 AU, where the tidal-disruption model favors such formation exterior to 
$\sim$ 2 AU, even though the starting column density of the fallback disk interior to 1 AU is no greater than that for the tidal 
disruption model.  On this basis, the supernova fallback model more plausibly explains the compact planetary 
system of PSR 1257+12.

There are several issues important to understanding the formation of pulsar planet 
systems that have been left untreated by this paper and should be the subject of future study.  
First, our models neglect the migration of solid grains after they 
have sedimented out in the disk.  Global migration of small grains in either a laminar (Youdin \& Shu 2002; 
Currie 2005) or turbulent (Stepinski \& Valageas 1997) can result in a systematic redistribution of solids in a disk, 
affecting the preferred locations for planetesimal and, eventually, planet formation.  While the gas disk in 
the supernova fallback model evolved on timescales much faster than the migration timescale for grains in the 
solar nebula (e.g. $10^{4}$ yr vs. $\sim$ $10^{5}$-$10^{6}$ yr from Youdin \& Shu), tracking the evolution of the 
solids (e.g. the method of Alexander \& Armitage (2007)) should be included to more accurately 
model the evolution of the solid body distribution once the disk cools enough to allow condensation.
Second, given the rarity of planets around millisecond pulsars, the formation of circumpulsar disks from 
which planets can form should be rather 
atypical.  While disks from the tidal-disruption scenario are obviously rare, the conditions for the formation 
of supernova fallback disks capable of forming planets should be investigated more thoroughly.
Third, more detailed pulsar timing data to constrain the nature of the fourth body will
 help to discriminate between the rival scenarios for forming pulsar planets.  The fallback disk 
scenario would predict that any fourth body at larger semimajor axes is small.  
A more massive body, $\sim$ 1 M$_{\oplus}$ or greater, would be more consistent with the 
tidal disruption model.

Finally, a major issue neglected in this paper is how planet formation 
would actually proceed in a circumpulsar disk.  The key issues with this 
are whether or not the final (post-oligarchic) stages of pulsar planet formation are 
characterized by growth (by mergers) like the terrestrial planets planets, how much material 
is needed to form the known planets, and how long the final assembly of planets takes to complete.
For the solar nebula, the final stages of planet formation interior to $\sim$ 2 AU
is primarily characterized by repeated mergers of sub-Earth mass 'oligarchs' (e.g. Kenyon \& Bromley 2006).  Whether or not 
the final stage of planet formation from supernova fallback disks is characterized by mergers or ejections 
of these lunar-mass oligarchs depends on escape velocity 
from their surfaces compared to the escape velocity from the system (Goldreich et al. 2004).  
The age of PSR 1257+12 also sets some limit on the timescale for planet formation.
The PSR 1257+12 system is probably 
$\sim$ $P/2\dot{P}$$\sim$ $8\times10^{8}$ years old (e.g. Miller \& Hamilton 2001).  
Therefore, the merger timescale must be less than $\sim 10^{9}$ years.  This timescale is somewhat larger than the 
timescale for merger and cleanup of bodies in the terrestrial zone of our solar system (Goldreich et al. 2004).  The merger timescale 
will depend critically on the amount of solid mass, $\Sigma_{d}$, the final planetary mass, $M_{p}$, and the orbital frequency.  
Numerical simulations of pulsar planet formation will help to determine the minimum $\Sigma_{d}$ allowable, and thus constrain 
the initial disk conditions from which these planets form.  We leave such simulations to future studies.  
\acknowledgements
We thank Randall Cooper, Phil Armitage, Charles Gammie, Alex Wolszczan, and Scott Kenyon for fruitful discussions 
and the anonymous referee for suggestions that improved the manuscript.  T. C. is partially supported by an SAO 
Predoctoral Fellowship.

\begin{deluxetable}{lllllll}
\tablecolumns{7}
\tablecaption{Orbital Parameters for the three confirmed planets in the PSR 1257+12 system}
\tablehead{{$Planet$} & {$M/(M_{\oplus})$}  & {$r$/1AU}   & {$Period$ (d)} & {$e$} &  {$i$ (both solutions)} & {$\omega$ (deg)}}
\startdata
$A$ & $0.02$ & $0.19$ & $25.262$ & $0$ & ... & $0.0$ \\
$B$ & $4.3$ & $0.36$ & $66.5419$ & $0.0186$ & $53, 127$ & $250.4$ \\
$C$ & $3.9$ & $0.46$ & $98.2114$ & $0.0252$ & $47, 133$ & $108.3$ \\
\enddata
\end{deluxetable}

\begin{deluxetable}{ccccc}
\tablecolumns{5}
\tablecaption{Frequency-Averaged Opacities}
\tablehead{{n} & {$\kappa_{o}$}   & {a} & {b} &  {$T_{max}(K)$}}
\startdata
$1$ &   $2.0\times10^{-4}$     &   0   &   2 &  167 \\
$2$ &   $2.0\times10^{16}$    &  0   & -7   &  203 \\
$3$ &   $1.0\times10^{-1}$   &  0    &  0.5   &  ... \\
$4$ &   $2.0\times10^{81}$   &  1   &  -24   &  ... \\
$5$ &   $1.0\times10^{-8}$   &  2/3   &  3   &  ... \\
$6$ &   $1.0\times10^{-36}$   &  1/3   &  10   &  ... \\
$7$ &  $1.5\times10^{20}$   &  1   &  -5/2   &  ... \\
$8$ &  $0.348$   &  0   &  0   & -  \\
\enddata
\tablecomments{Opacity regimes in order of increasing temperature according to the $\kappa = \kappa_{o}\rho^{a}T^{b}$
power law.  The fits are from Bell \& Lin(1994). 
The maximum temperatures for each regime $n$, which for $n \geq3$ depends on gas density, are found by
matching the values for $\kappa_{n}$ and $\kappa_{n+1}$.  The opacity prescription in the n=4-5 regimes is modified as discussed in 
\S 3.1.}
\end{deluxetable}

\clearpage

\begin{figure}
\plottwo{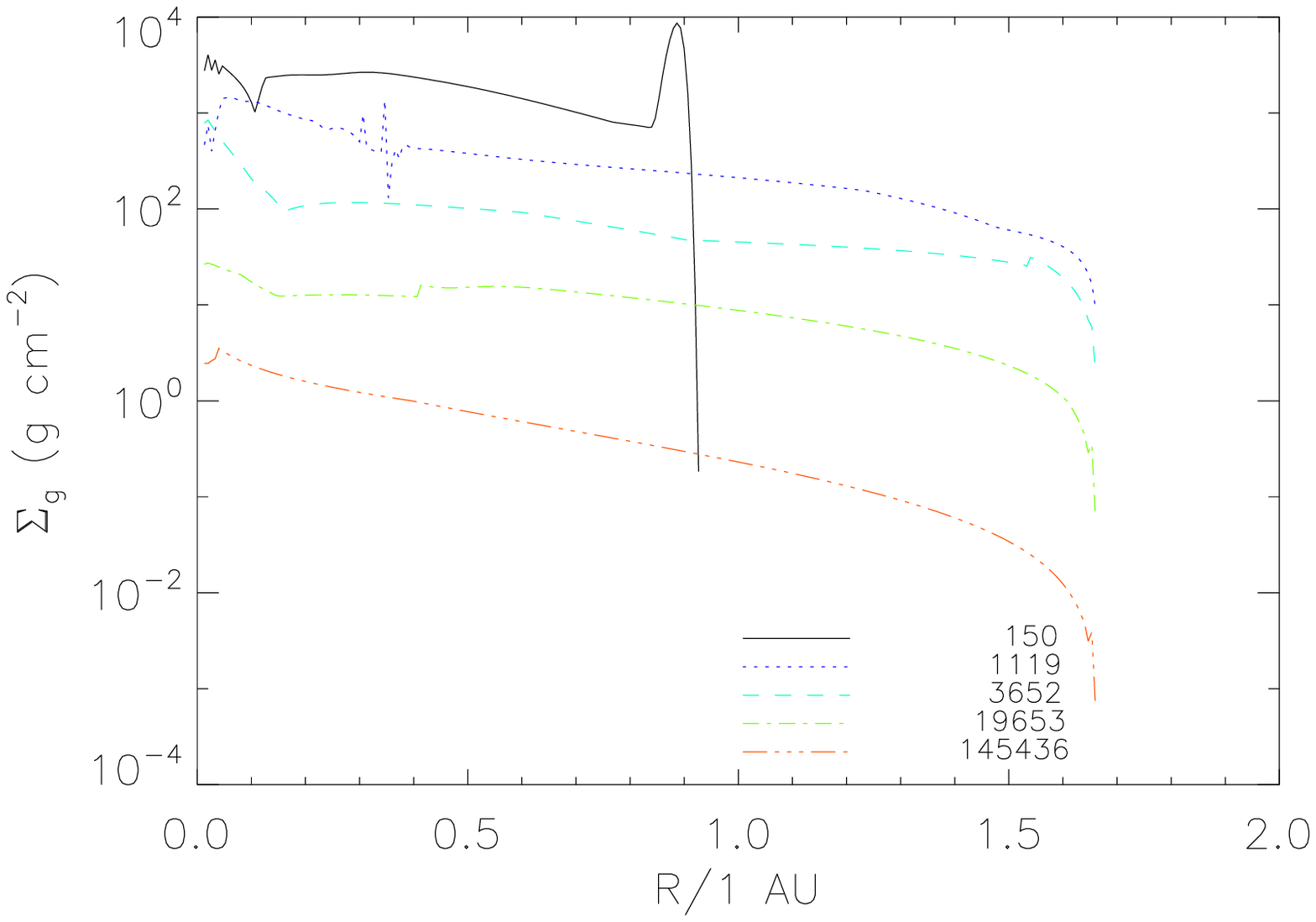}{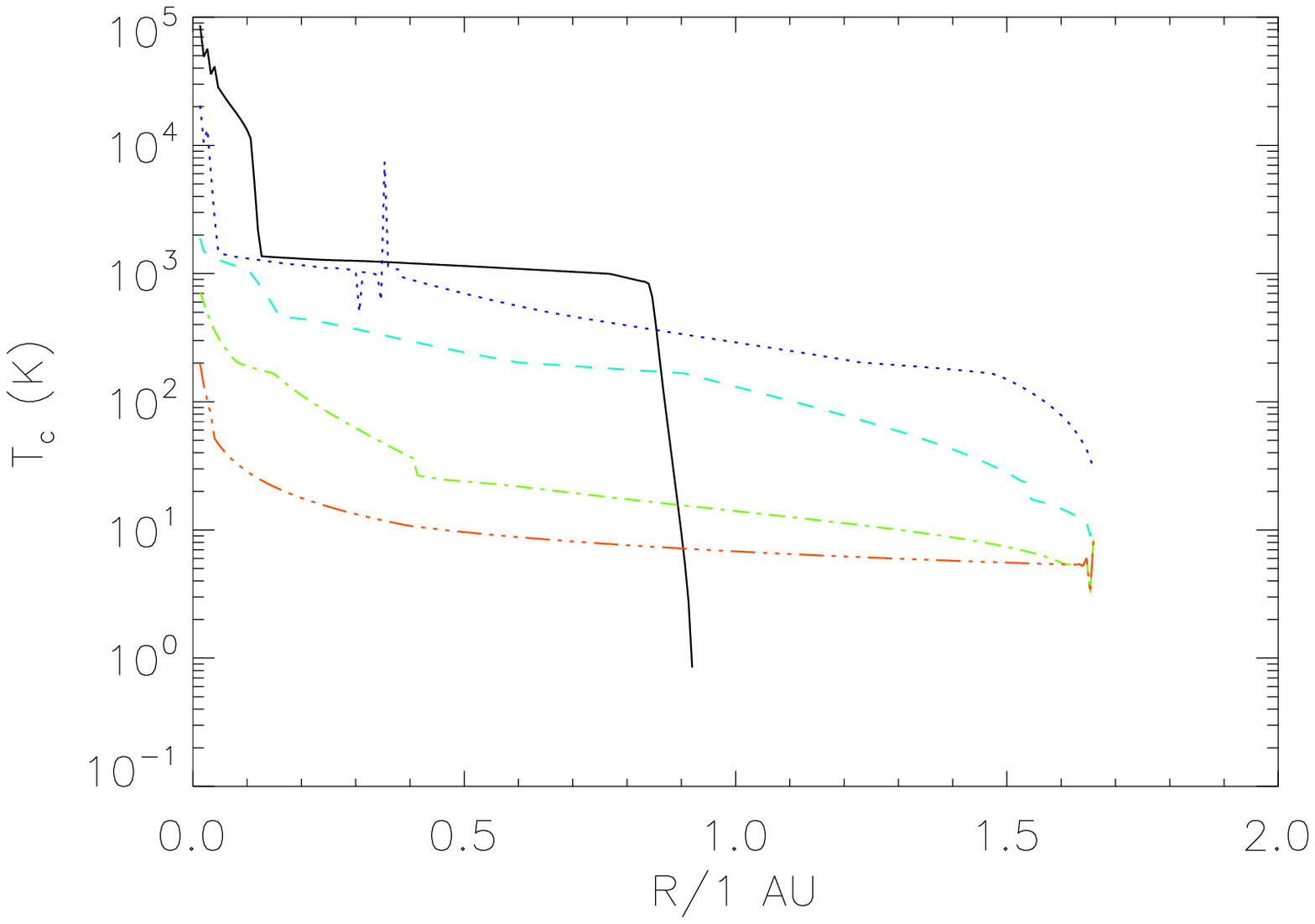}
\caption{Evolution of the gas surface density ($\Sigma_{g}$) and midplane temperature ($T_{c}$) 
profiles for the supernova-fallback disk model, assuming fully viscous accretion.  The evolution 
timescale appears to be $\tau$ $\sim$ $10^{4}$ years, much shorter than typical 
circumstellar disk evolution timescales ($\sim 10^{6}-10^{7}$ years).} 
\end{figure}
\begin{figure}
\plottwo{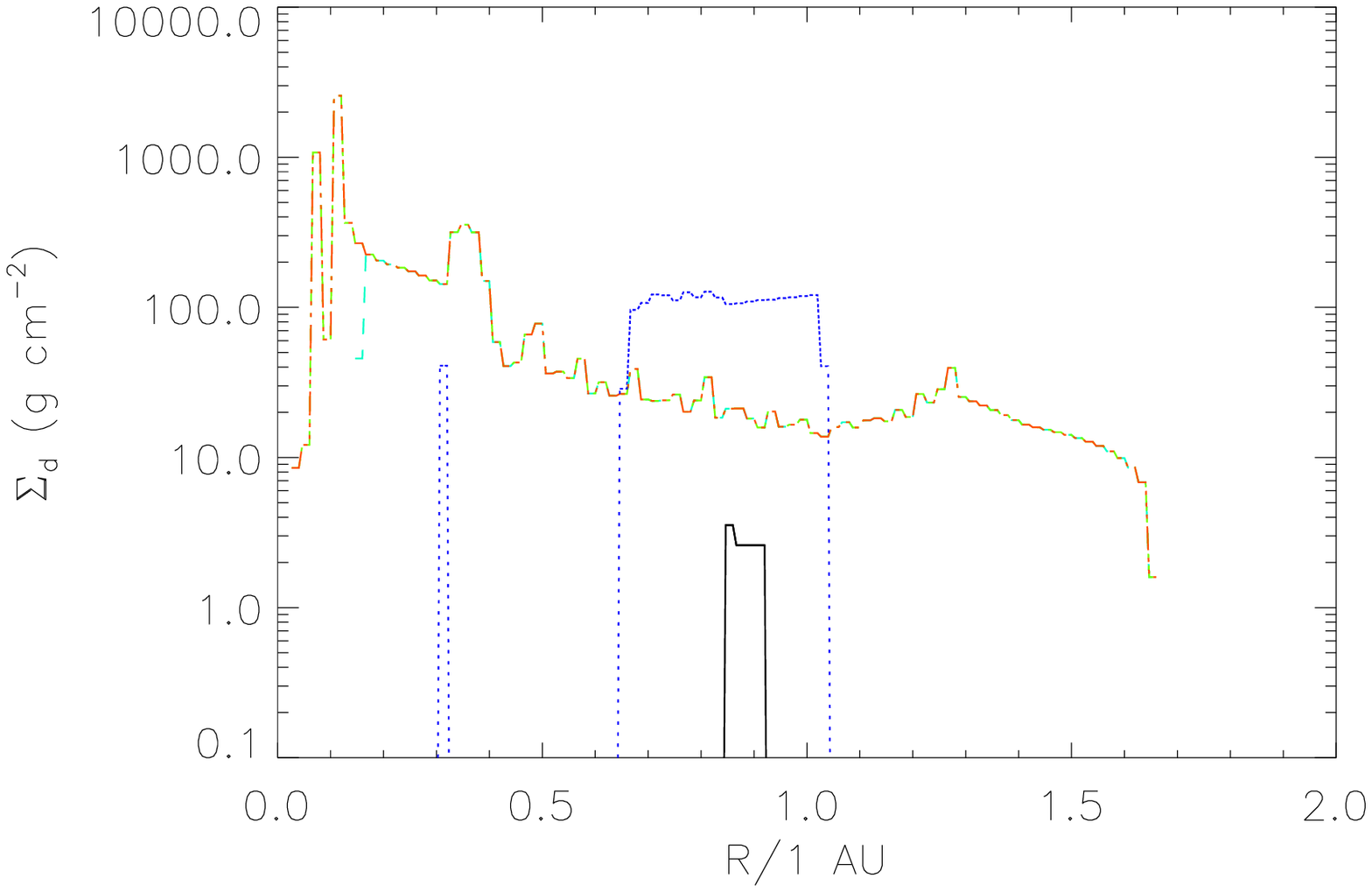}{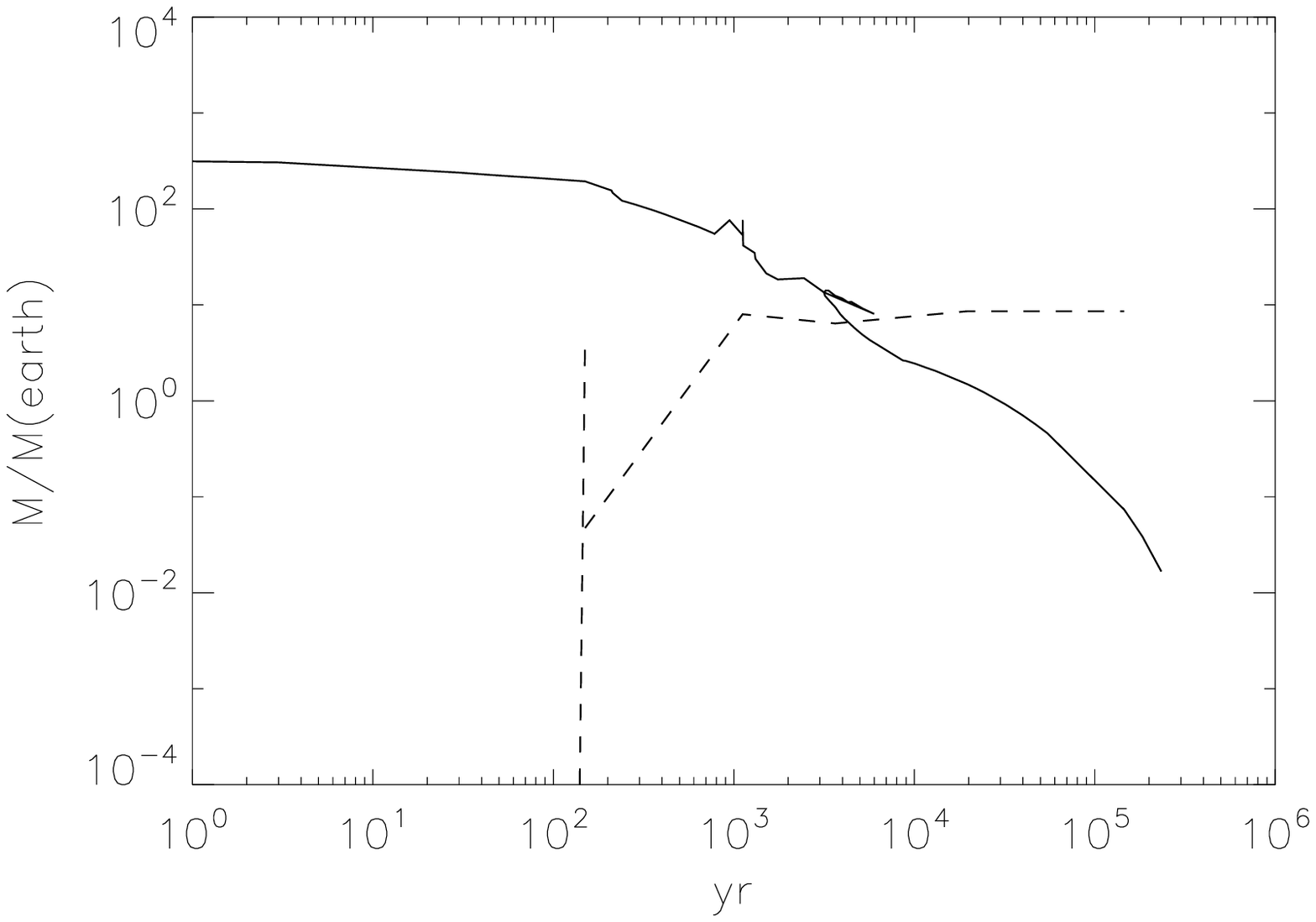}
\caption{(left)Evolution of the solid body surface density ($\Sigma_{d}$) for 
the supernova-fallback disk model, assuming fully viscous accretion.  
Typical $\Sigma_{d}$ are $\sim$ 10-100$\times$ Minimum Mass Solar Nebula (MMSN) values from 0.1-0.5 AU.  
(right)Mass interior to 1 AU for the supernova-fallback disk model, assuming
fully viscous accretion.  The gas mass (solid line) drops below 1 $M_{\oplus}$ by $\sim$ 10$^{4}$ years.  
From $t\sim10^{3}$ yr and later, several earth masses of solids 
(dotted line) reside interior to 1 AU, comparable to the total mass of the three known planets
of PSR 1257+12.  
}
\end{figure}

\begin{figure}
\plottwo{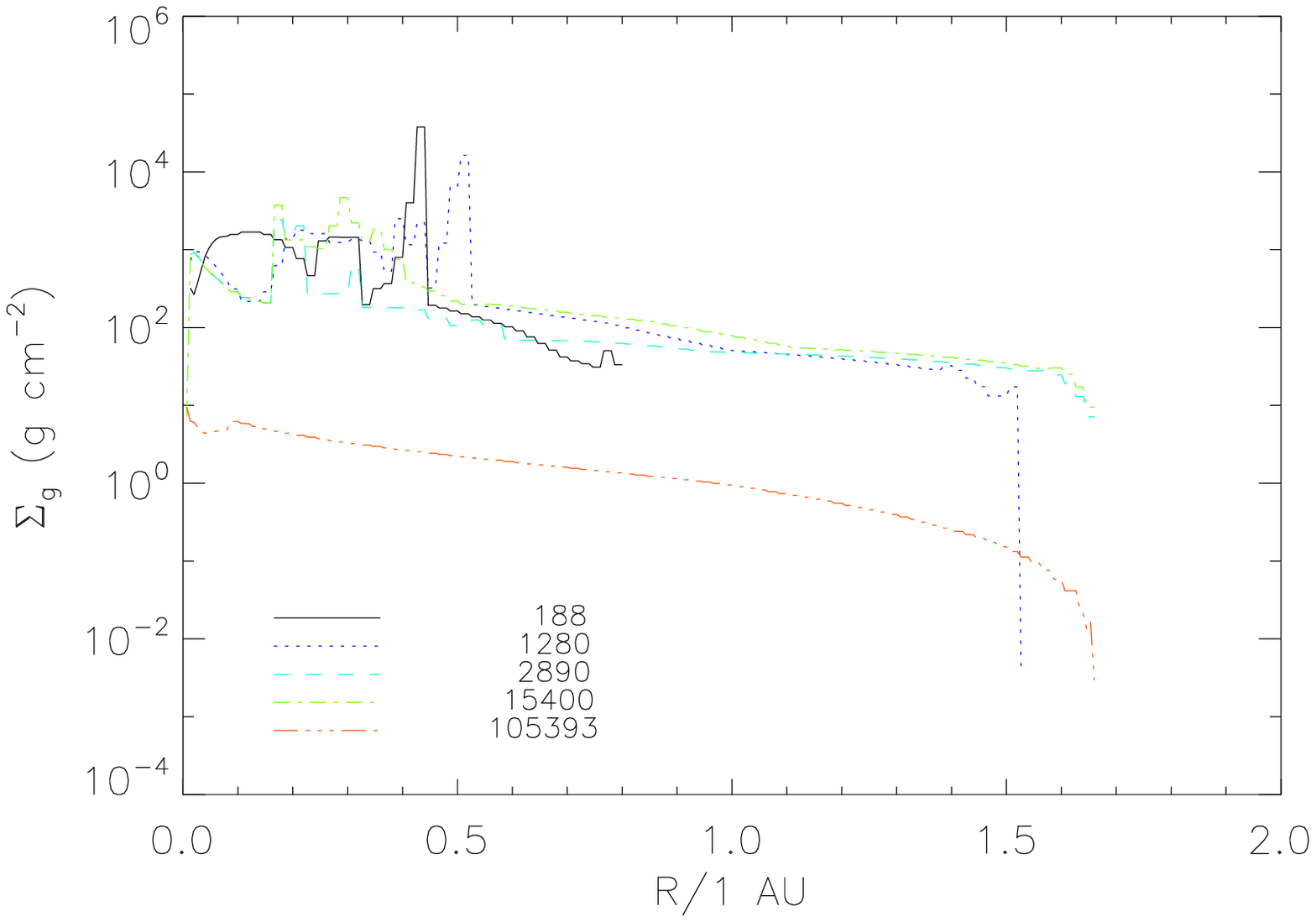}{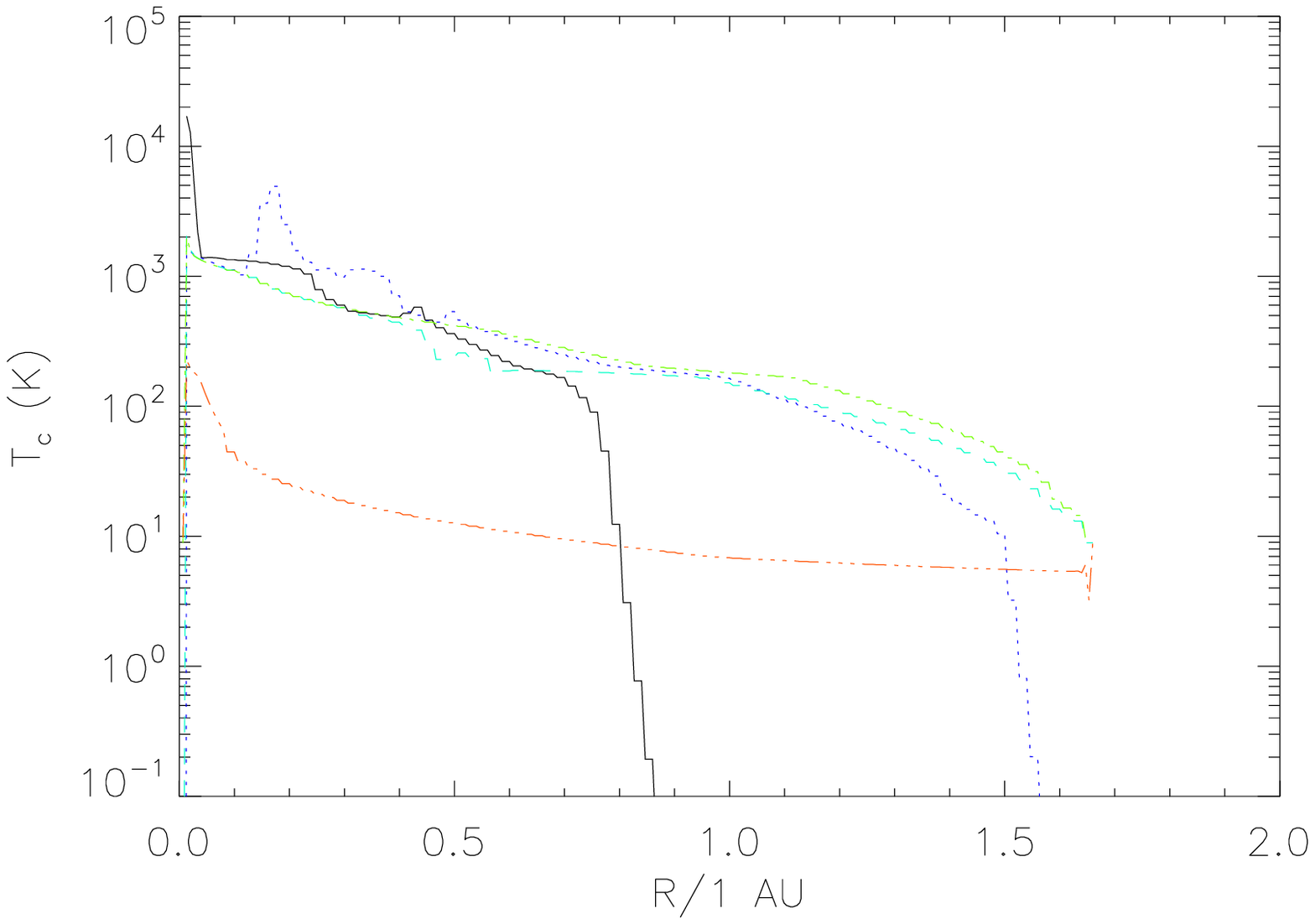}
\caption{Evolution of the gas surface density ($\Sigma_{g}$) and midplane temperature ($T_{c}$) 
profiles for the supernova-fallback disk model, assuming layered accretion.  The evolution 
timescale appears to be slightly longer than the fully viscous case, $\tau$ $\sim$ $10^{5}$.} 
\end{figure}
\begin{figure}
\plottwo{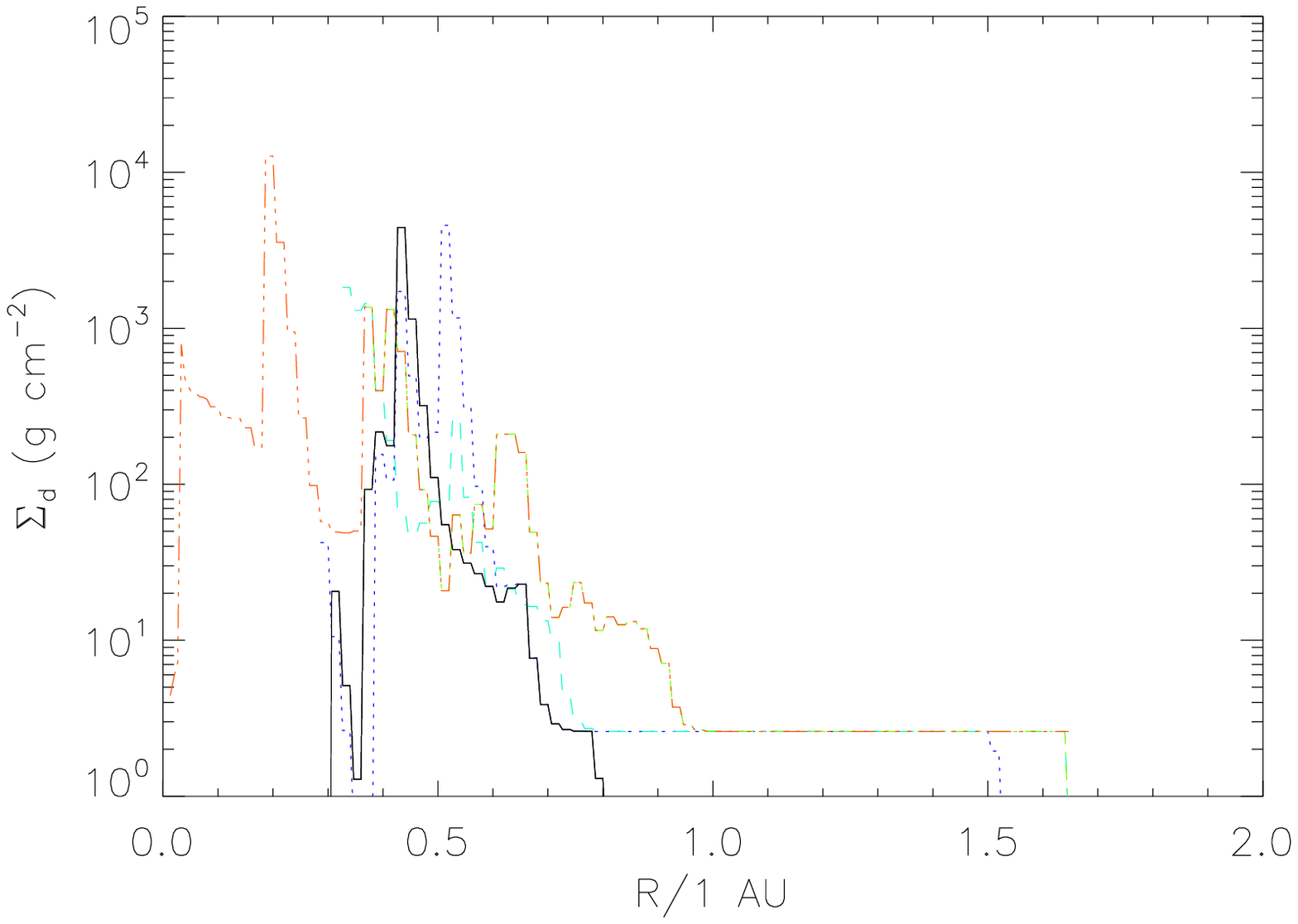}{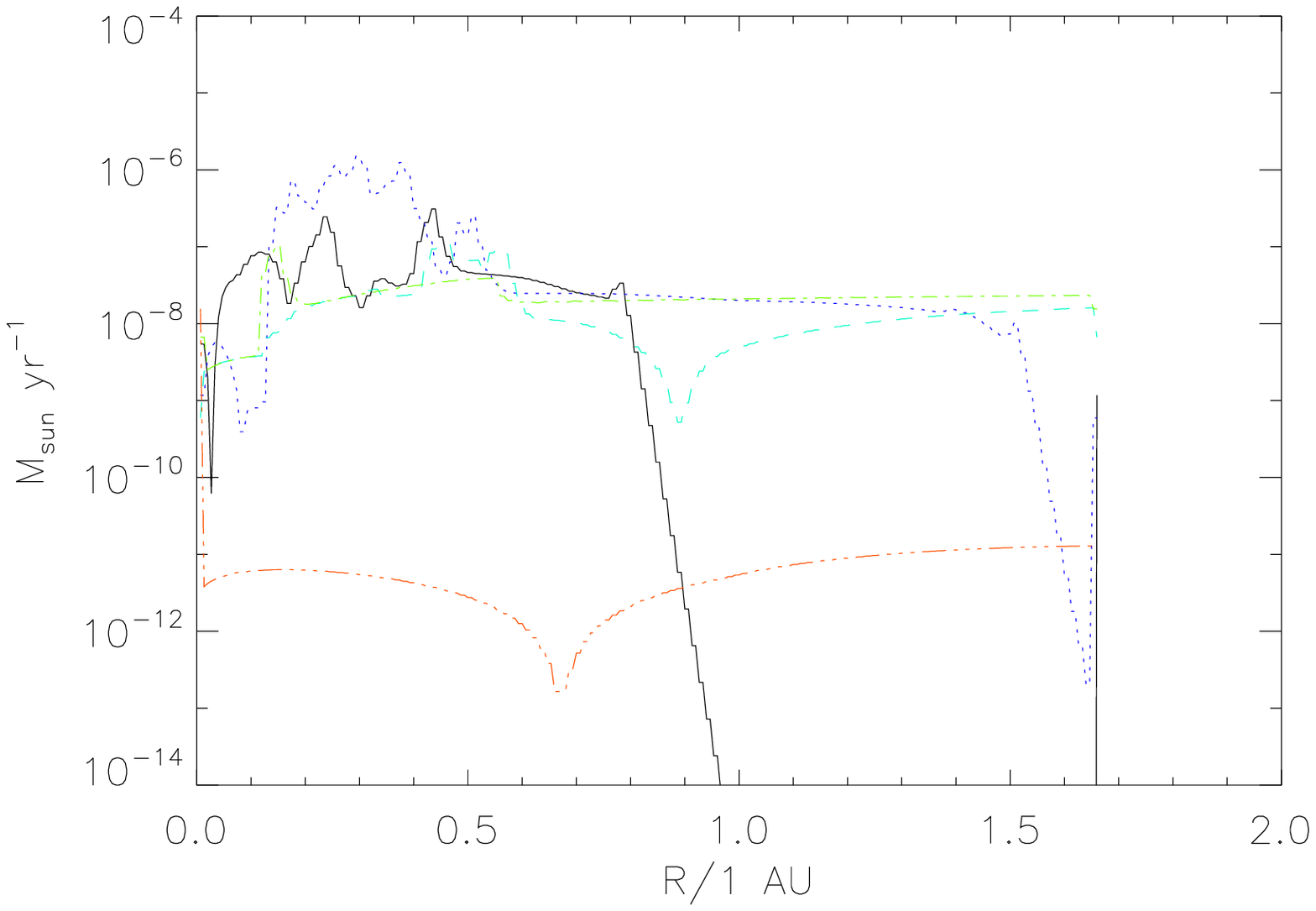}
\caption{Evolution of the solid body surface density ($\Sigma_{d}$) and accretion rate profiles for 
the supernova-fallback disk model, assuming layered accretion.  A high mass of solids is deposited 
near the present semimajor axes of the known pulsar planets.  The accretion rate shown 
is the absolute value of the accretion rate: the local minimums reached (e.g. at 0.7 AU for 
t=$\sim$10$^{5}$ years) occur because the accretion rate changes sign from positive to negative.
}   
\end{figure}
\begin{figure}
\plottwo{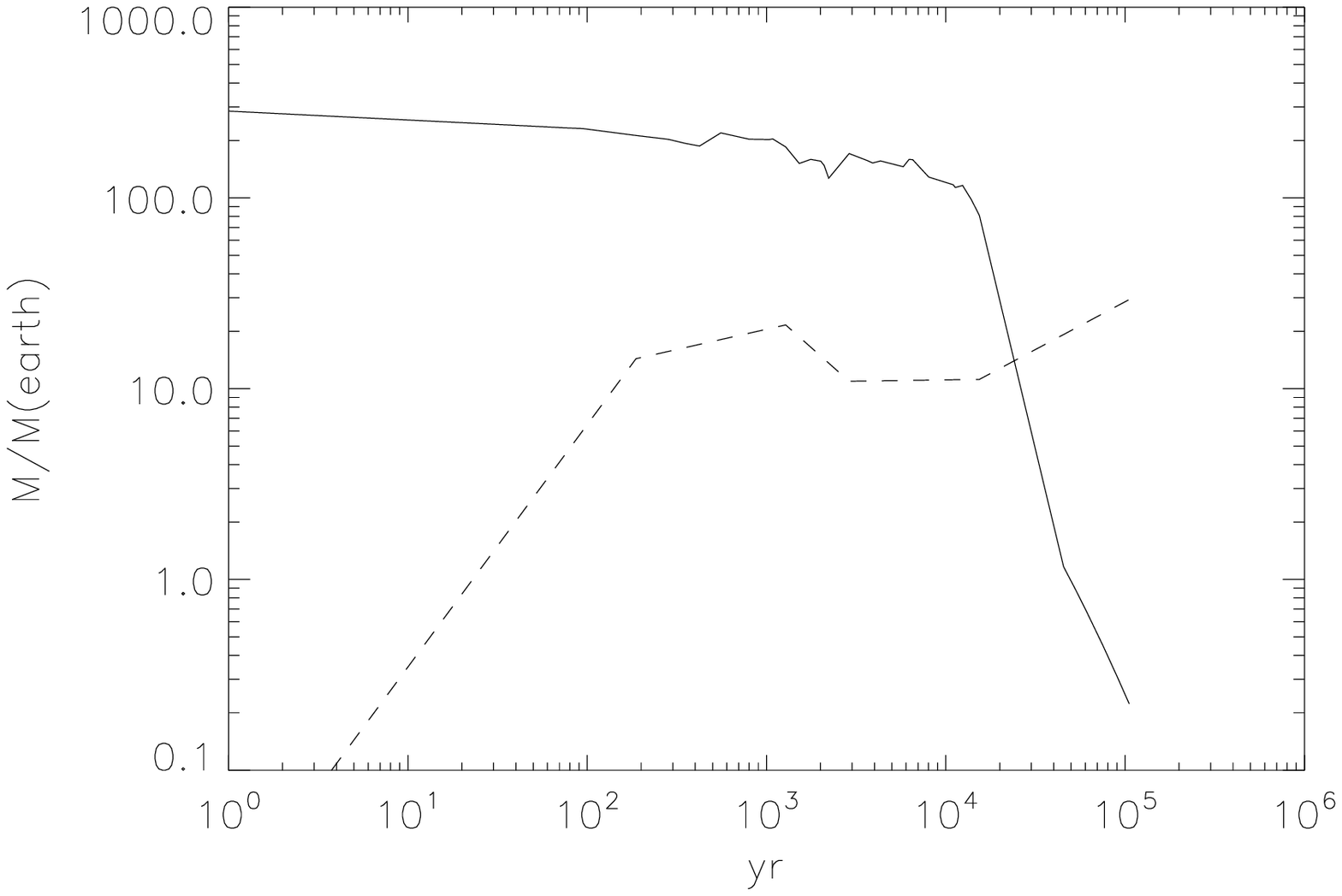}{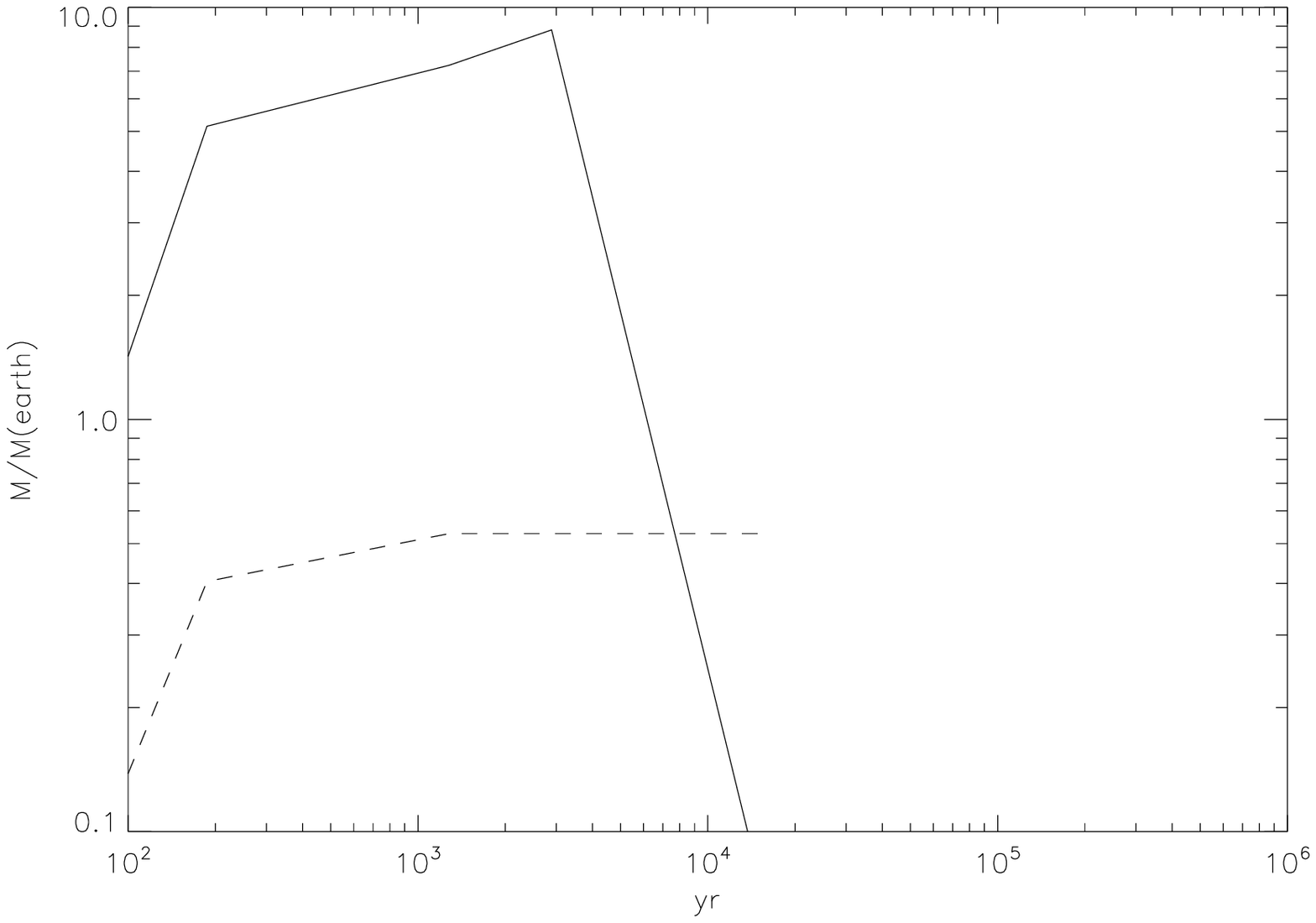}
\caption{Mass interior to 1 AU and exterior to 2 AU for the supernova-fallback disk model, assuming 
layered accretion.  The supernova-fallback model deposits up to $\sim$ 30 $M_{\oplus}$ of solid material 
within 1 AU and $\sim$ 0.1-0.5 $M_{\oplus}$ beyond 2 AU.  The likely result is the formation of 
one or more $\gtrsim$ 1 $M_{\oplus}$ bodies interior to 1 AU and perhaps Mars or smaller-sized bodies 
beyond $\sim$ 2 AU.  The formation of gas giant planets is highly unlikely, owing to the fast gas 
dissipation timescale.} 
\end{figure}
\begin{figure}
\plottwo{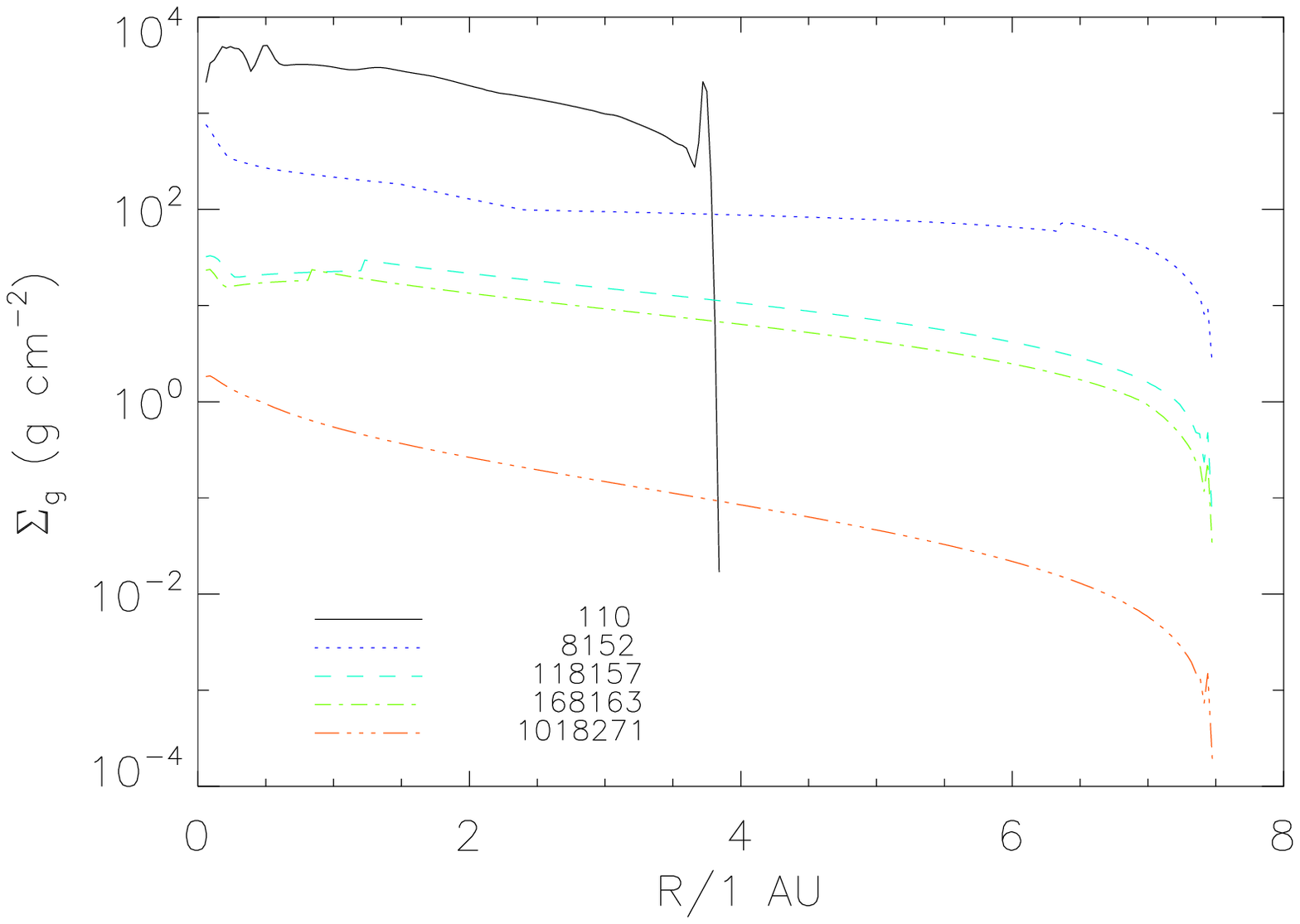}{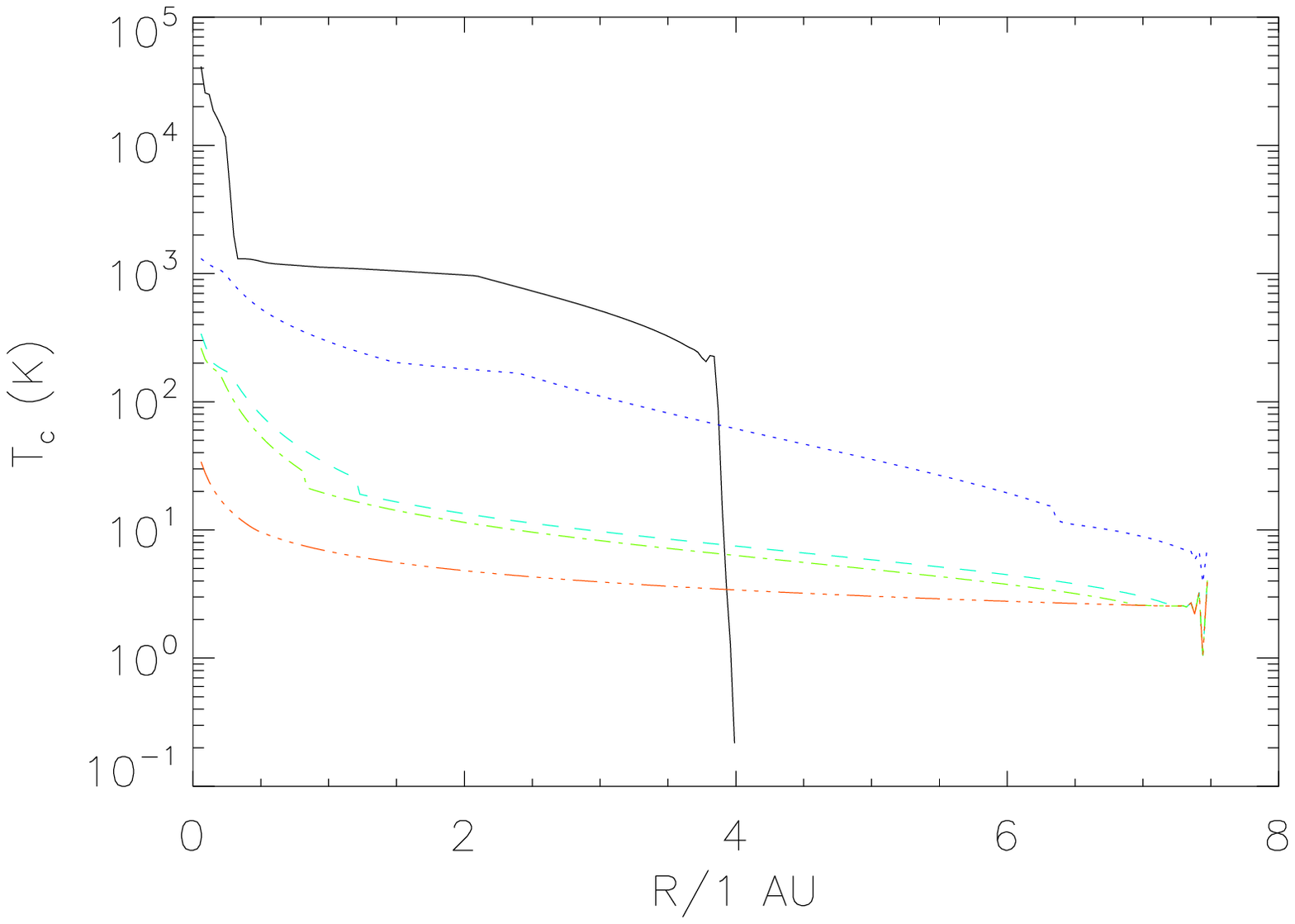}
\caption{Evolution of the gas surface density ($\Sigma_{g}$) and central temperature profiles for 
the tidal-disruption disk model, assuming fully viscous accretion and initial angular momentum of 
$J\sim 10^{51}$ ergs$\cdot$s.  The disk evolves on $\tau \sim 10^{4}$ year timescales.} 
\end{figure}
\begin{figure}
\plotone{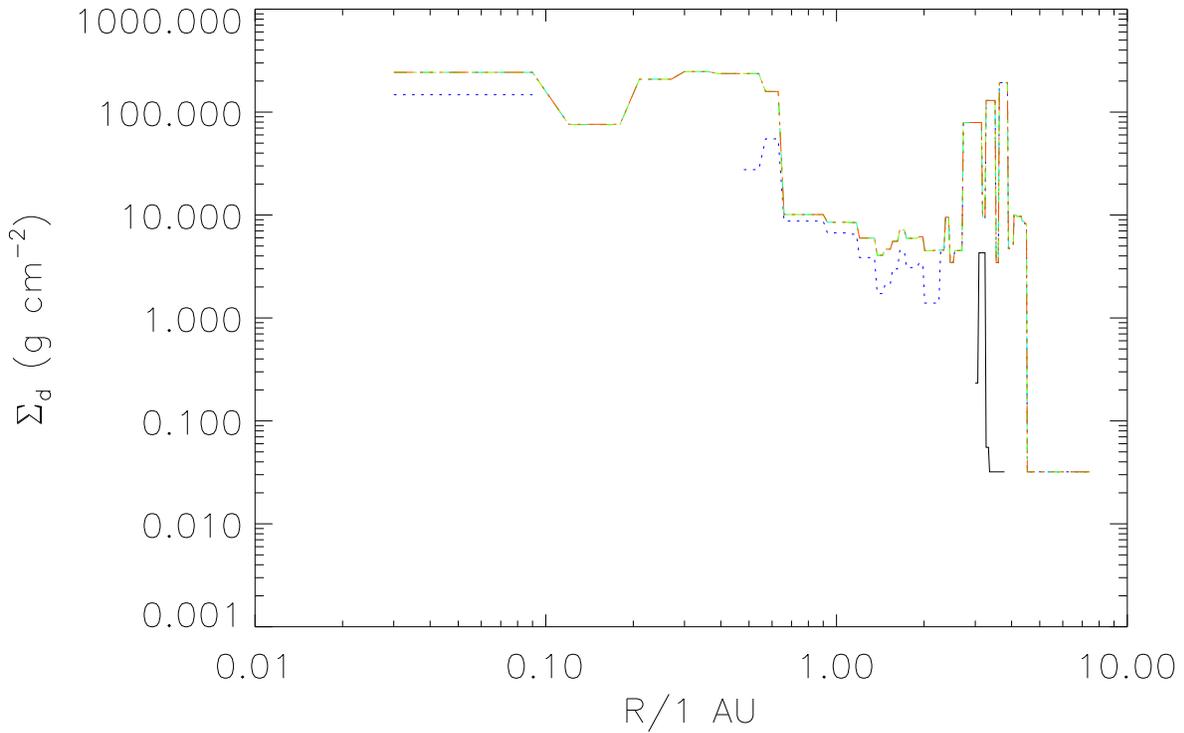}
\caption{Evolution of the solid surface density ($\Sigma_{d}$) and accretion rate profiles for 
the tidal-disruption disk model, assuming fully viscous accretion and initial angular momentum of 
$J\sim 10^{51}$ ergs$\cdot$s.  Between $\sim$ 2 and 4 AU the disk attains near MMSN values but drops 
sharply in solid content beyond $\sim$ 5 AU.  
} 
\end{figure}
\begin{figure}
\plottwo{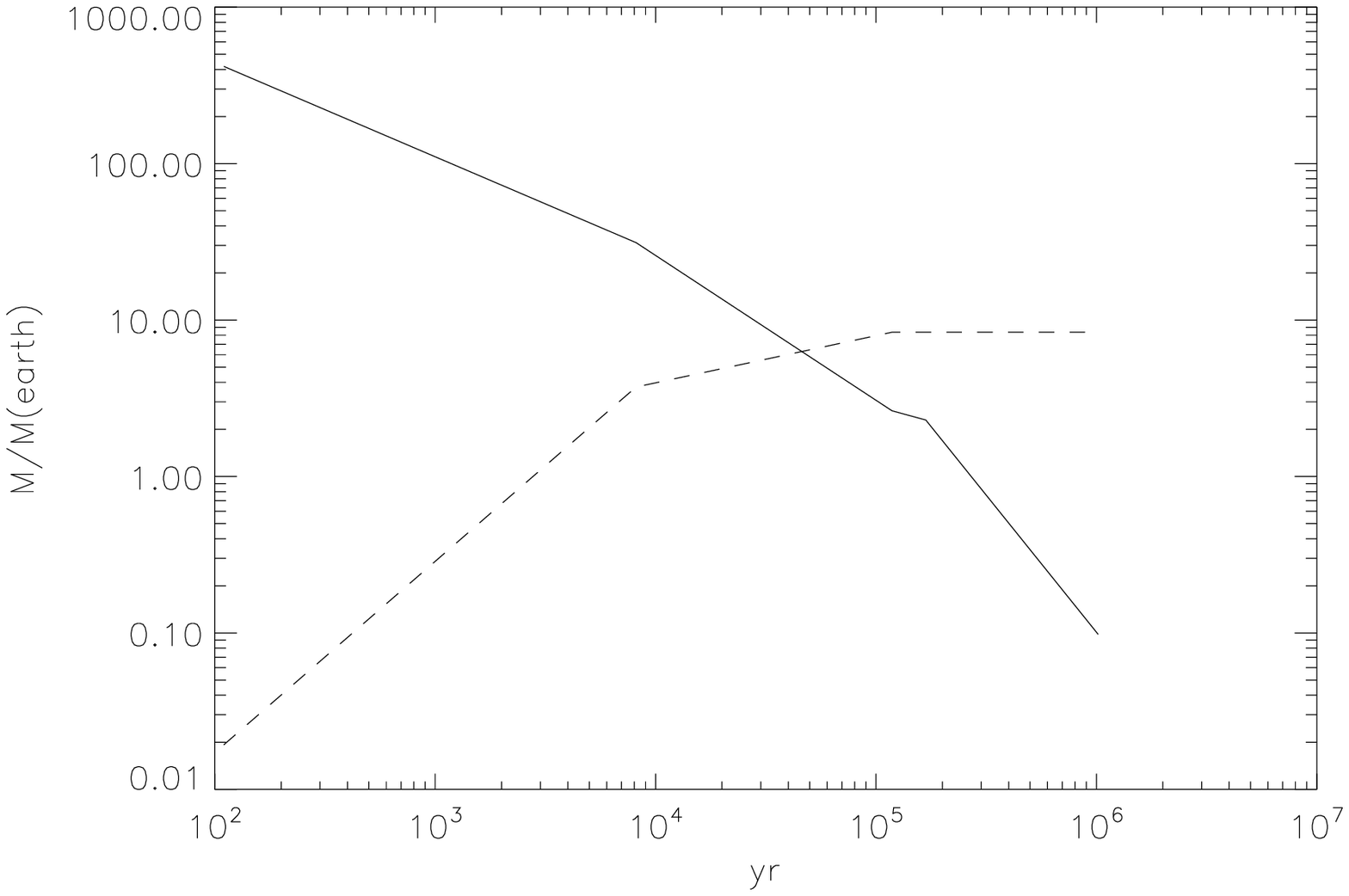}{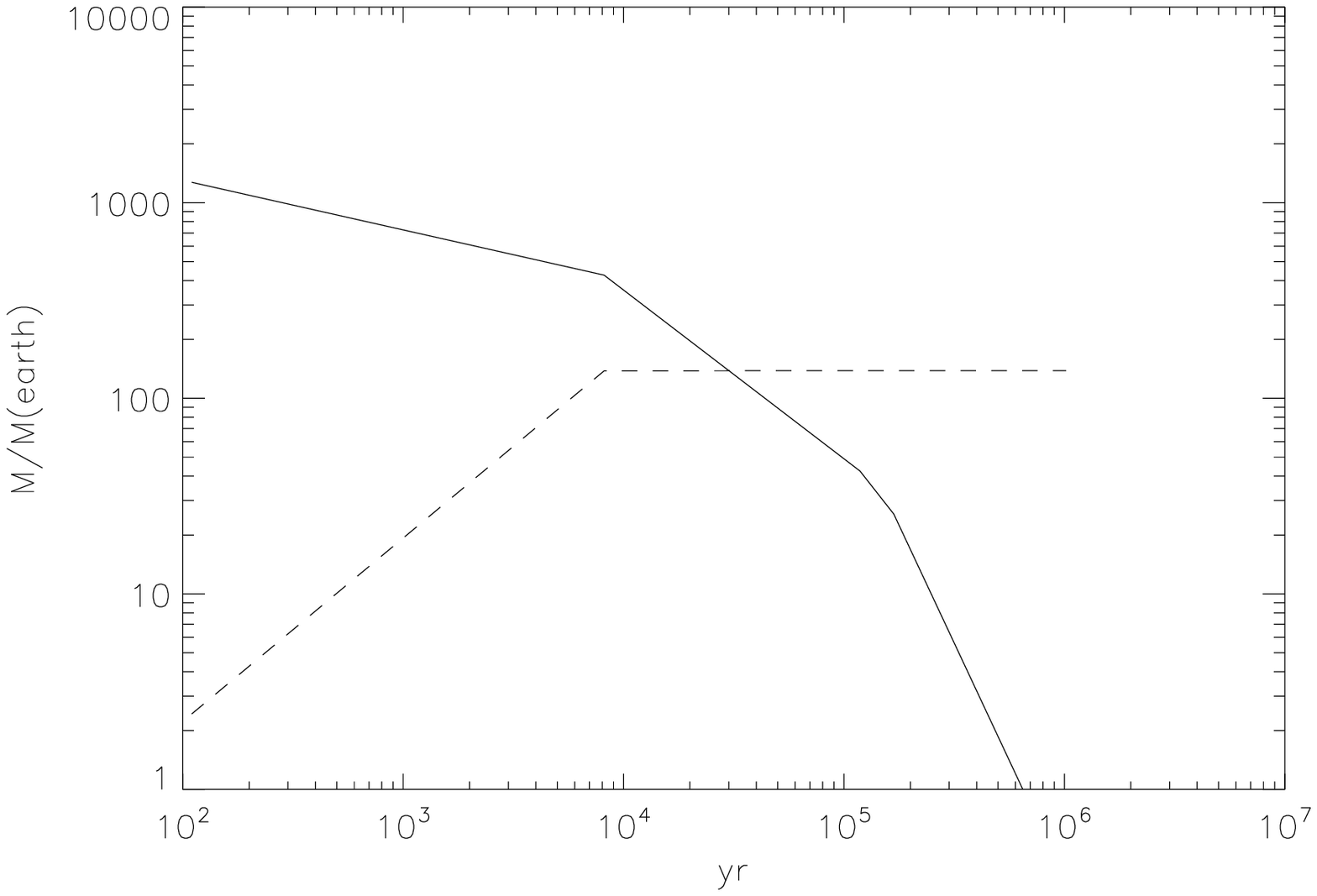}
\caption{The disk mass interior to 1 AU and exterior to 2 AU for 
the tidal-disruption disk model, assuming fully viscous accretion and initial angular momentum of 
$J\sim 10^{51}$ ergs$\cdot$s.  The gas mass $<$ 1 AU drops below 1$M_{\oplus}$ by $\sim 2\times10^{5}$ years 
from an initial value of $\gtrsim 100 M_{\oplus}$.  The mass beyond 2 AU declines on $10^{4}-10^{5}$ year timescales.
} 
\end{figure}
\begin{figure}
\plottwo{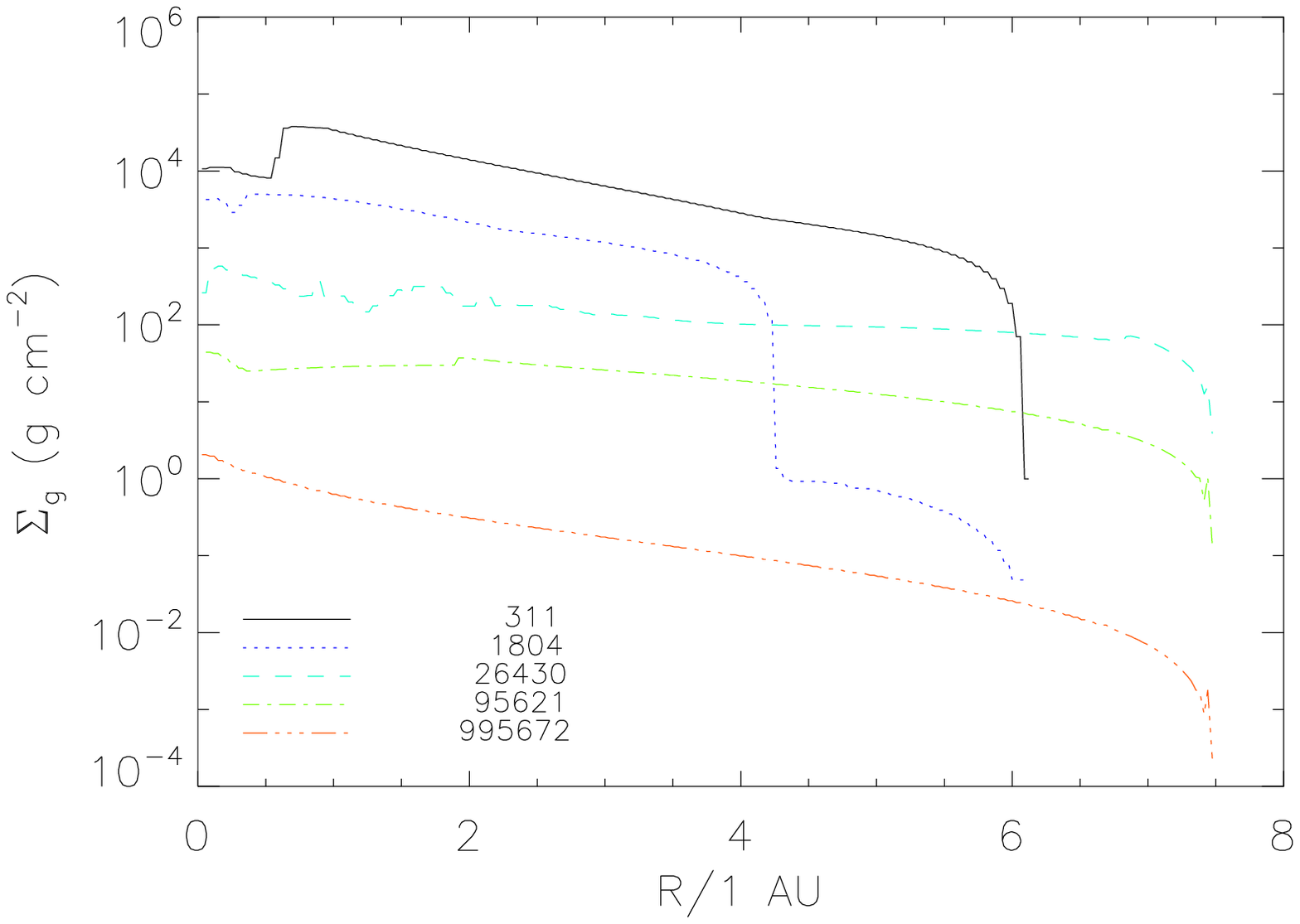}{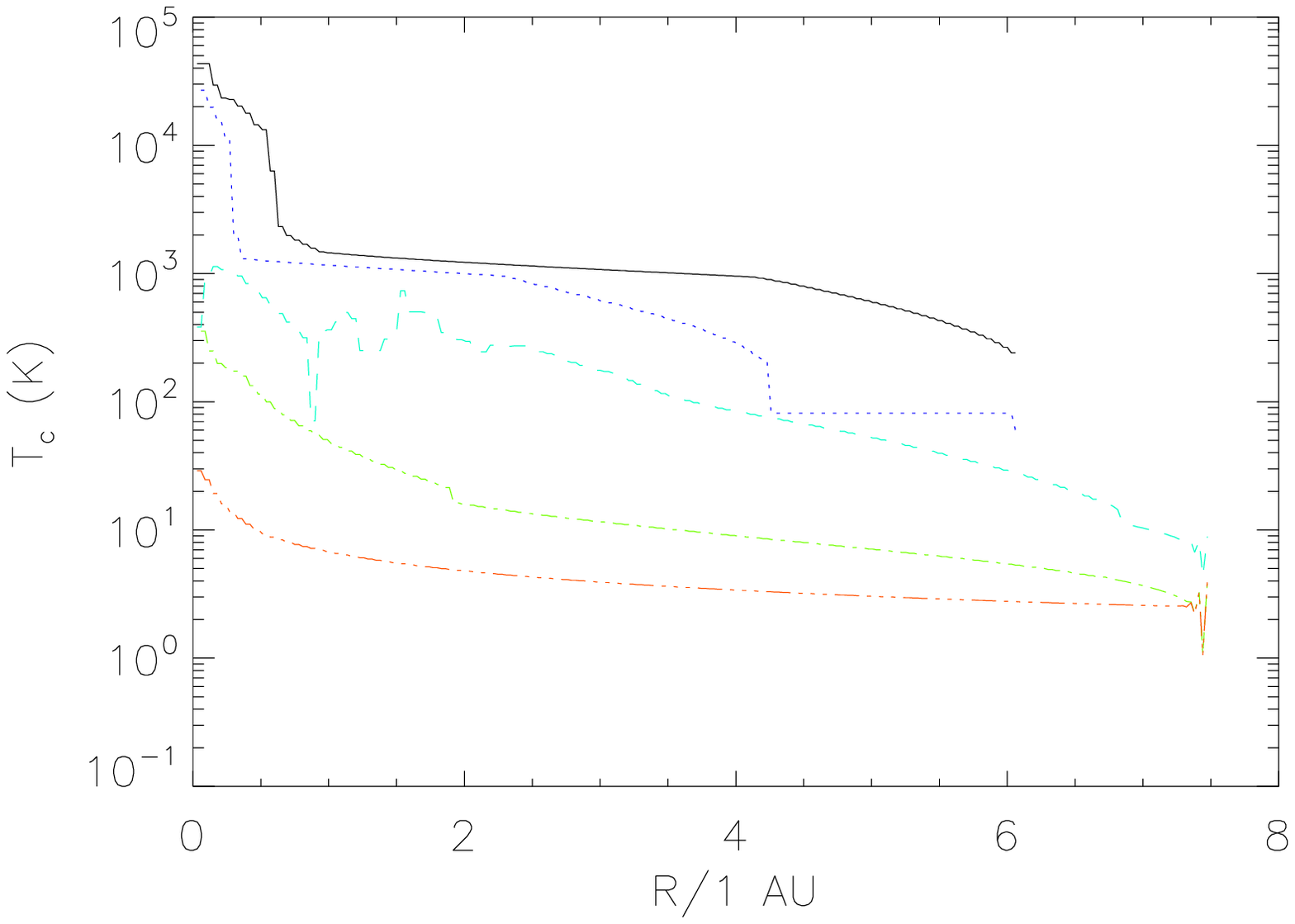}
\caption{Evolution of the gas surface density ($\Sigma_{g}$) and central temperature profiles for 
the tidal-disruption disk model, assuming fully viscous accretion and an initial angular momentum of 
$J\sim 10^{52}$ ergs$\cdot$s.  The disk evolution is slightly longer but similar to that for the $J\sim 10^{51}$ model, 
despite the order-of-magnitude increase in mass.} 
\end{figure}
\begin{figure}
\plotone{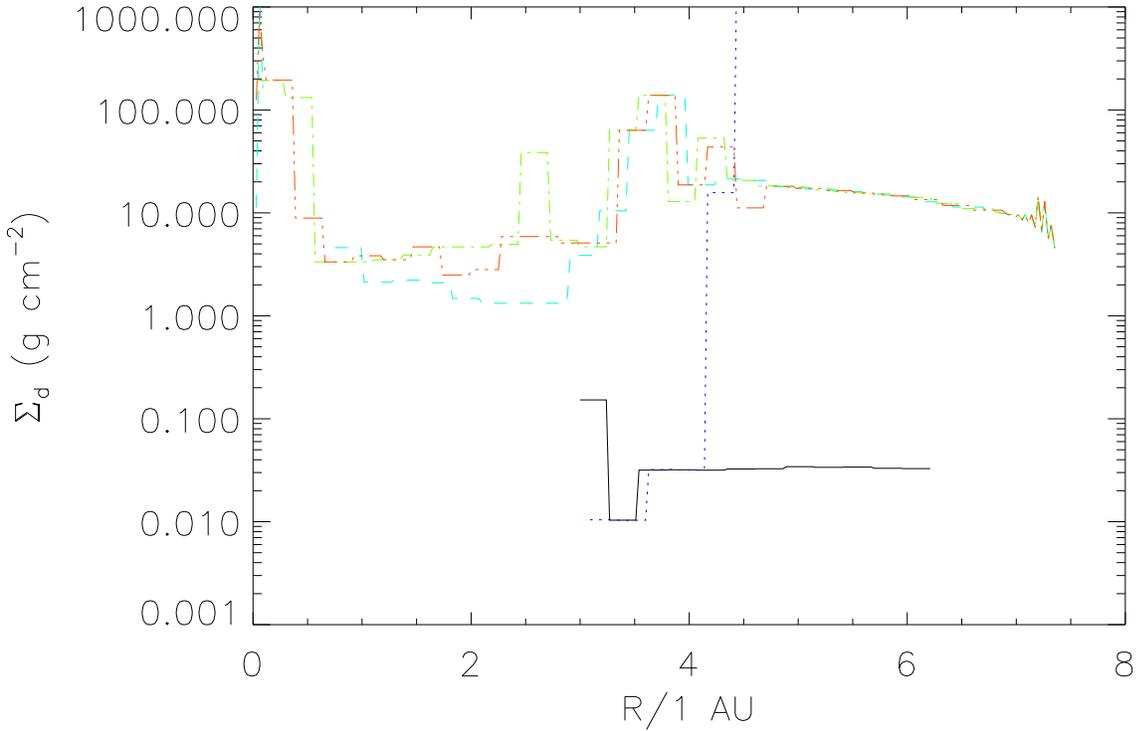}
\caption{Evolution of the solid surface density ($\Sigma_{d}$) for
the tidal-disruption disk model, assuming fully viscous accretion and initial angular momentum of
$J\sim 10^{52}$ ergs$\cdot$s.  $\Sigma_{d}$ is slightly larger than MMSN through $\sim$ 8 AU.}  

\end{figure}
\begin{figure}
\plottwo{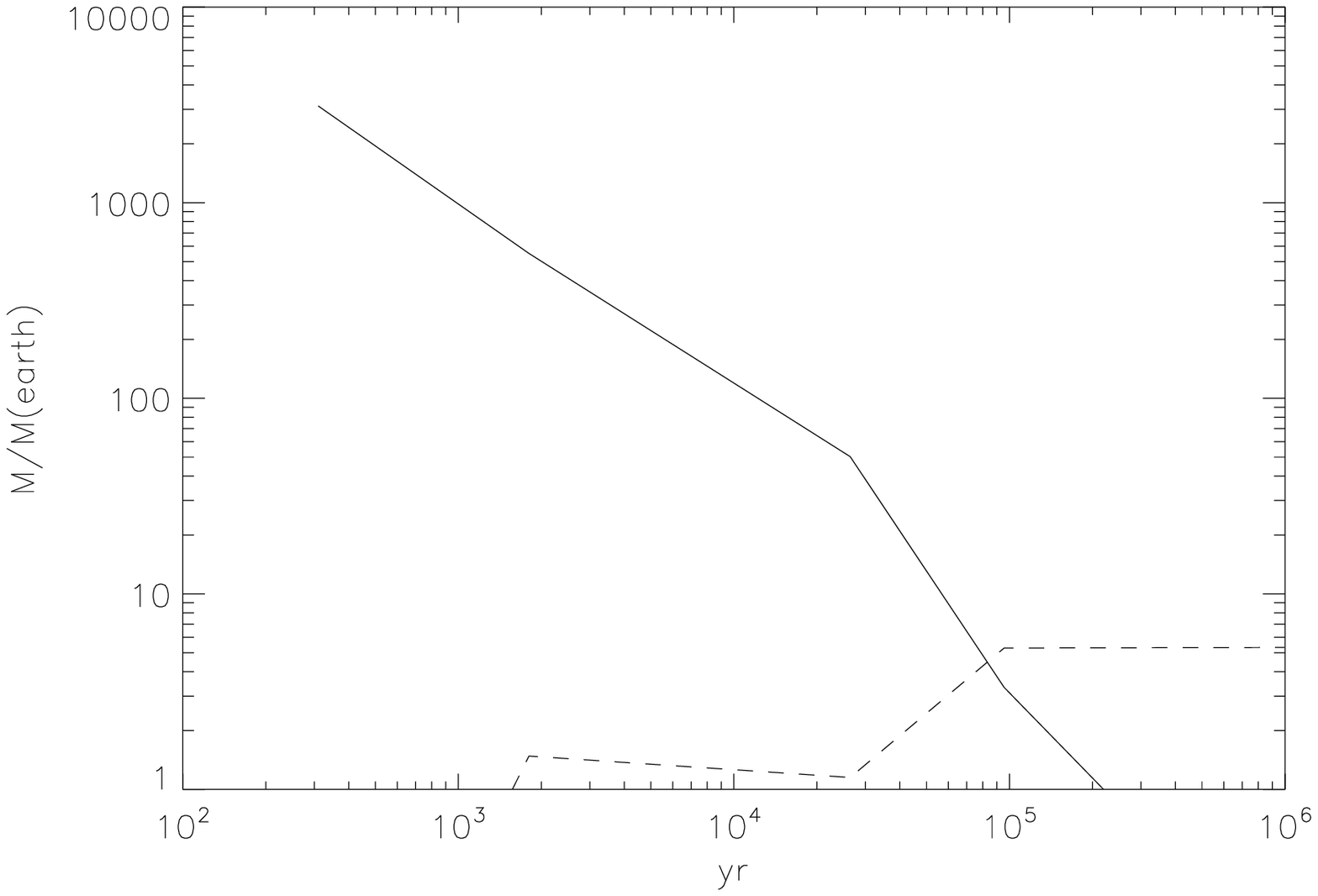}{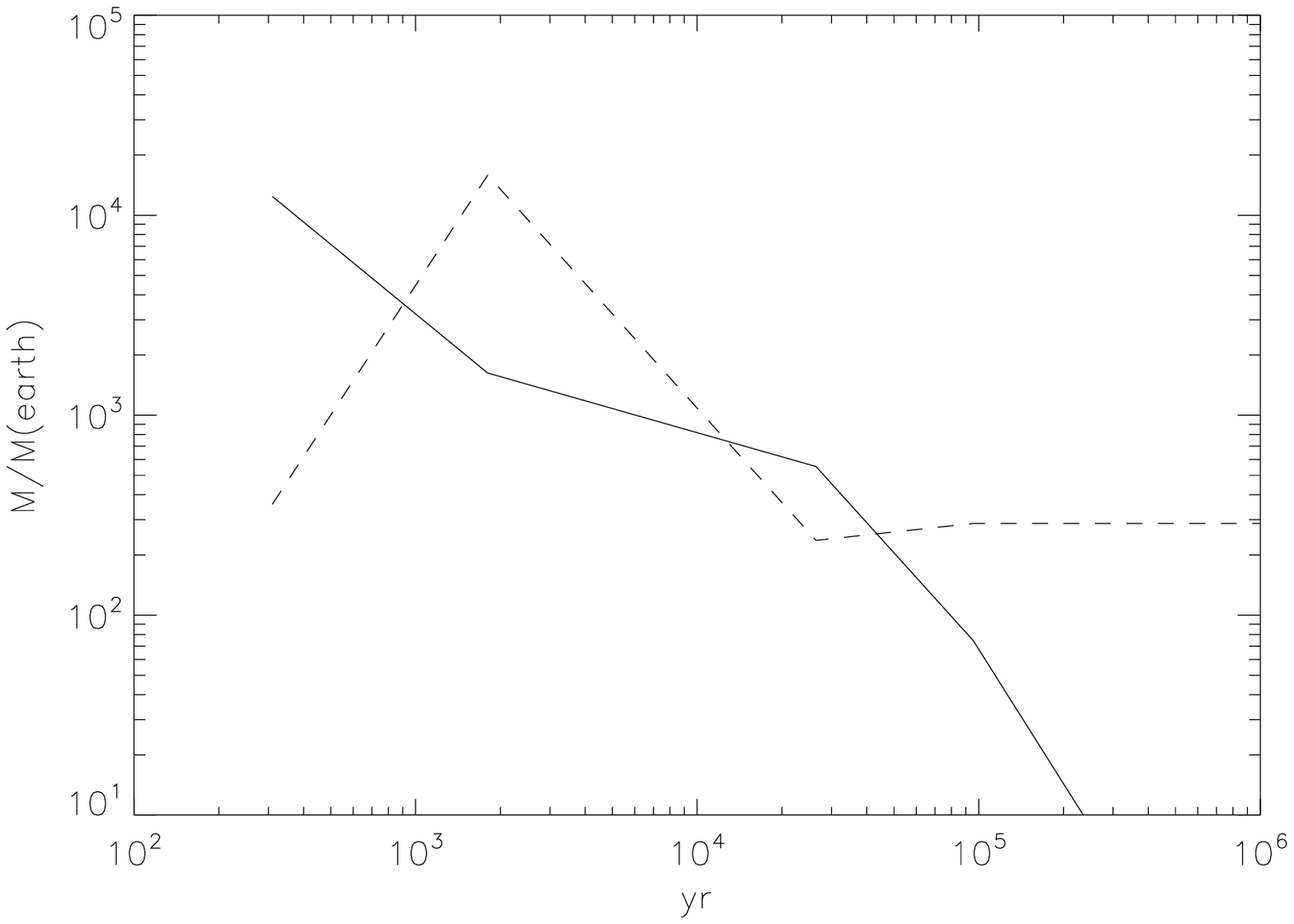}
\caption{The disk mass interior to 1 AU and exterior to 2 AU for
the tidal-disruption disk model, assuming fully viscous accretion and initial angular momentum of
$J\sim 10^{52}$ ergs$\cdot$s.  The inner disk does not cool sufficiently to deposit solids until its mass has significantly 
dropped.  The outer ($\ge$ 2 AU) regions of the disk cools fast initially and then heats up again as more material is 
transported into the outer disk regions.  About 200 $M_{\oplus}$ of solids are 
deposited by $\sim 10^{5}$ years.  However, the gas mass drops to $\sim 10 M_{\oplus}$ by $10^{5}$ years, making 
gas giant planet formation highly unlikely.}
\end{figure}
\begin{figure}
\plottwo{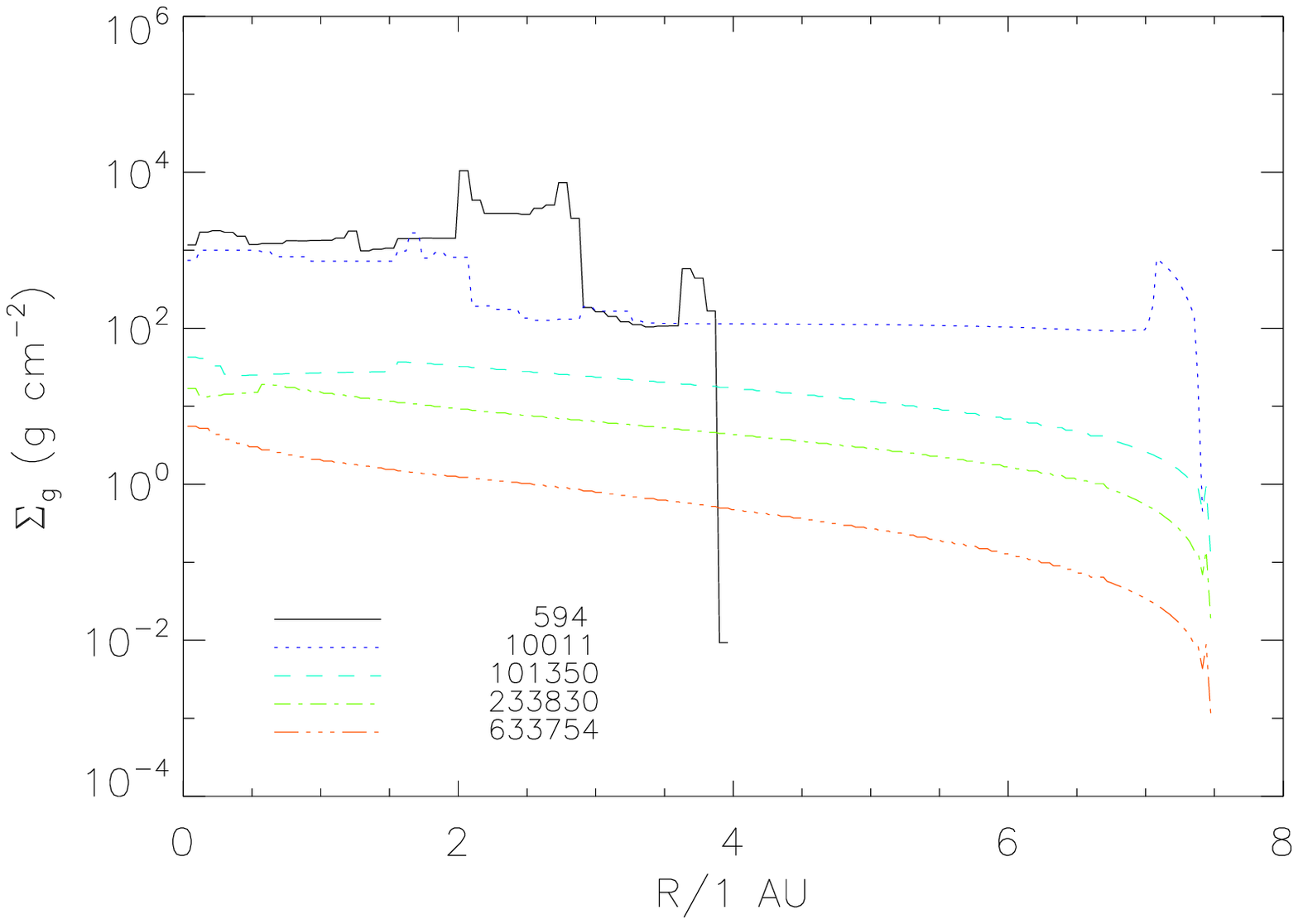}{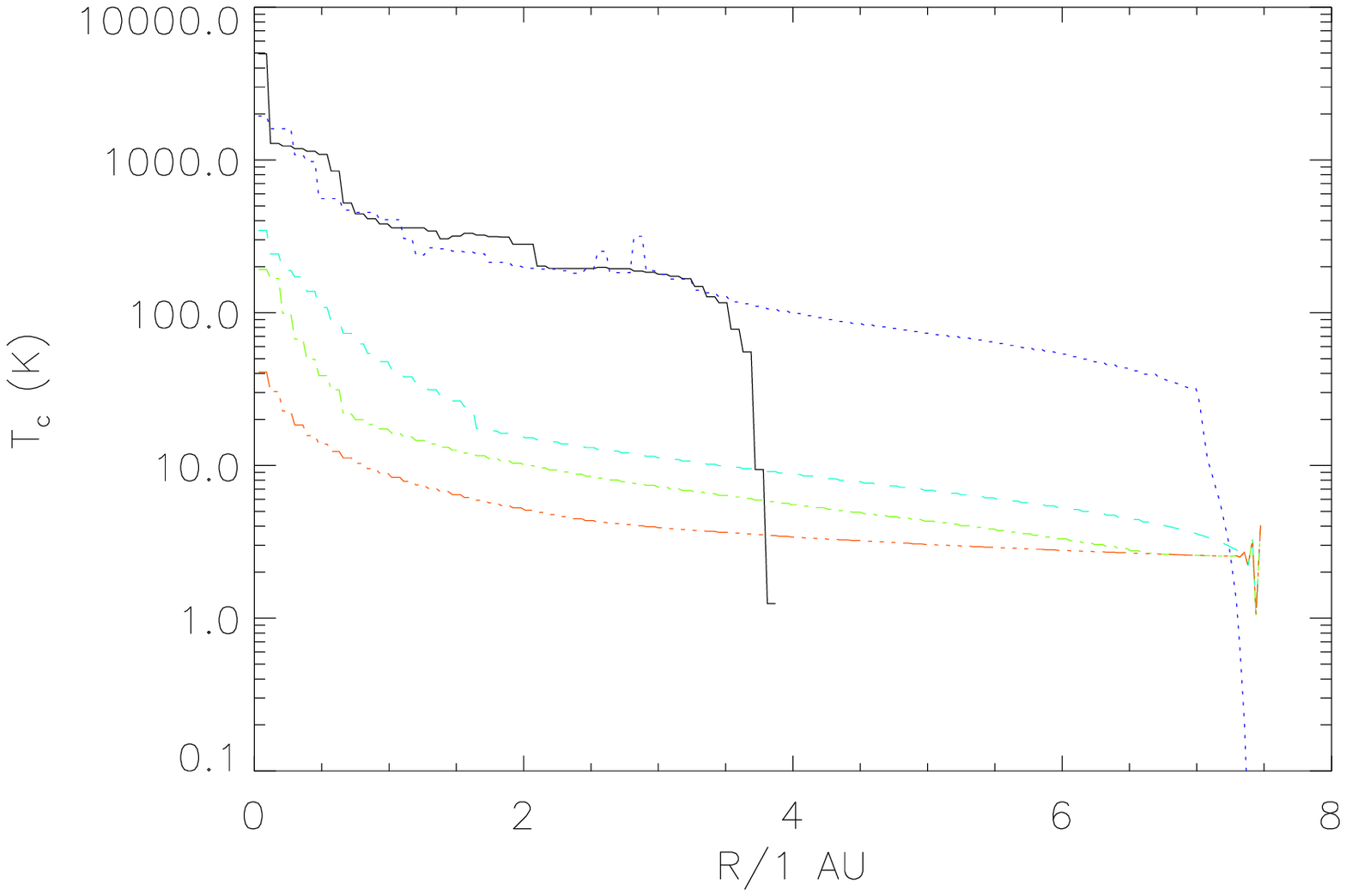}
\caption{Evolution of the gas surface density ($\Sigma_{g}$) and central temperature profiles for 
the tidal-disruption disk model, assuming layered accretion and an initial angular momentum of 
$J\sim 10^{51}$ ergs$\cdot$s.  The disk assumes a layered state for $t \lesssim 10^{5}$ years and the 
overall evolution timescale is longer than for the fully viscous models and all supernova-fallback models.} 
\end{figure}
\begin{figure}
\plottwo{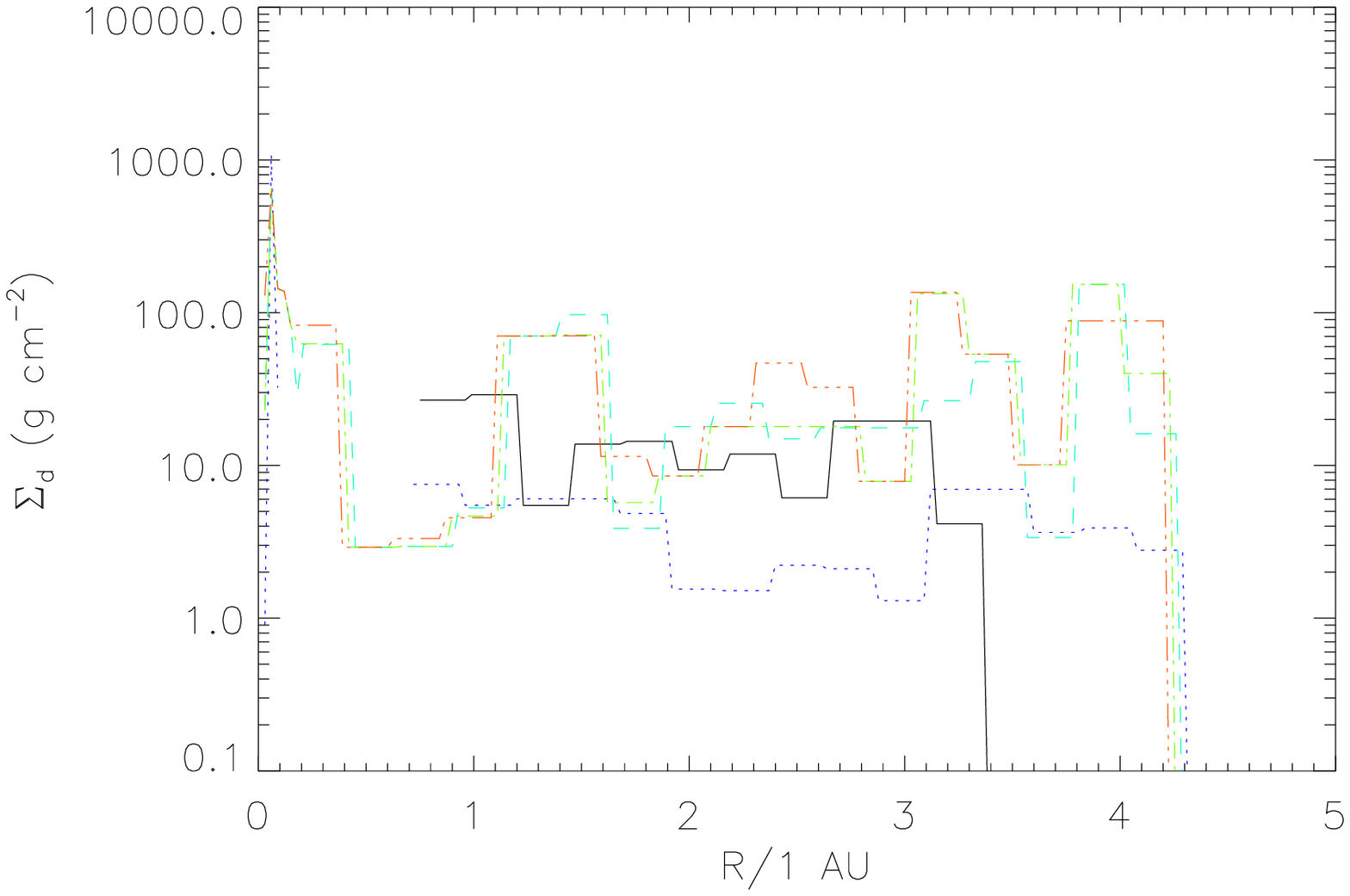}{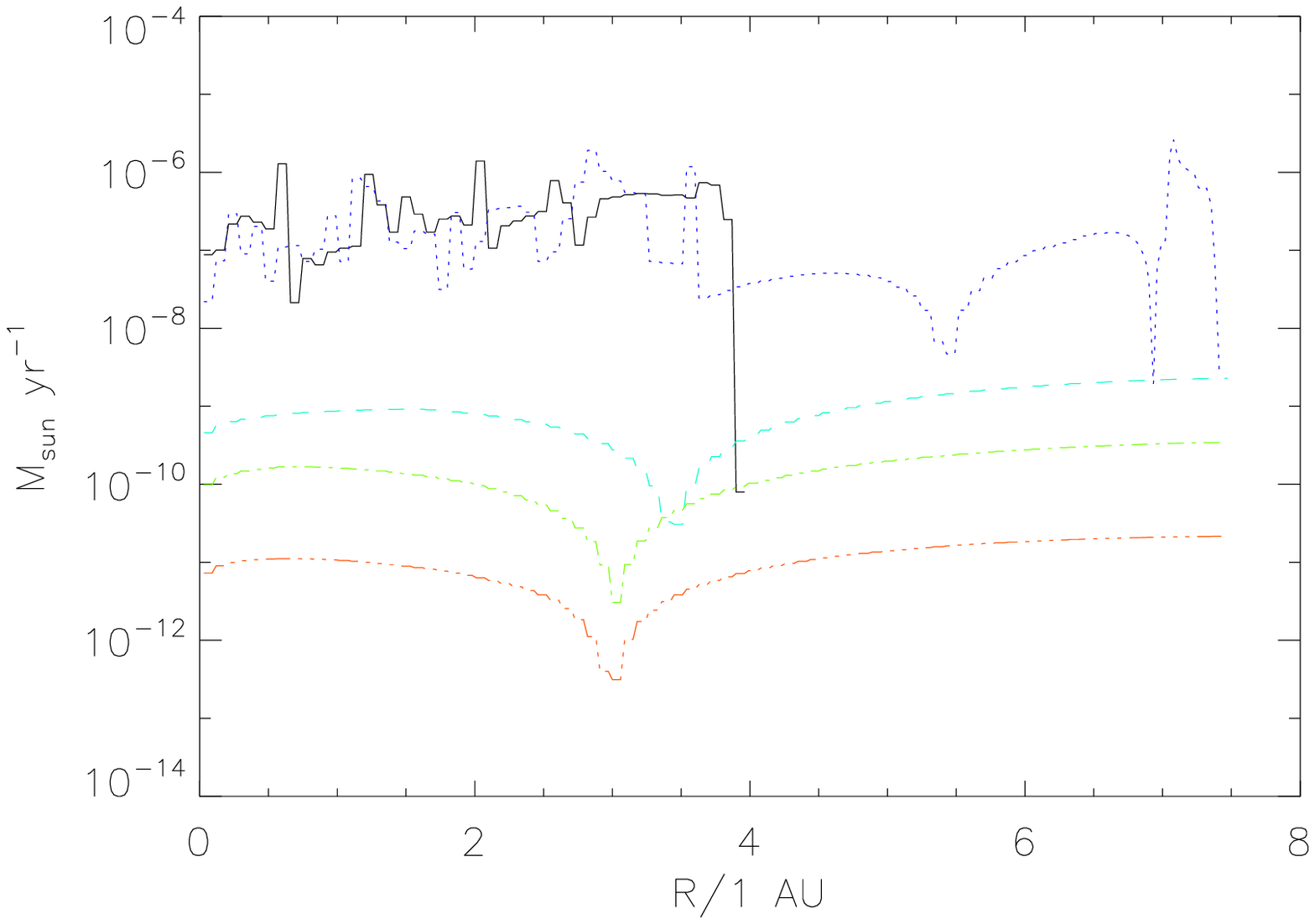}
\caption{Evolution of the solid surface density ($\Sigma_{d}$) and accretion rate profiles for
the tidal-disruption disk model, assuming layered accretion and initial angular momentum of
$J\sim 10^{51}$ ergs$\cdot$s.  A substantial solid column density forms by $\sim$ 10$^{5}$ years 
and fluctuates between 10 and 100 g cm$^{-2}$, with a median value of $\sim$ 30-40 g cm$^{-2}$.
During the layered accretion state the disk typically has a flat or 
inverted mass accretion rate profile throughout the disk, consistent with that expected for a layered, 
episodically accreting disk (Armitage, Livio, \& Pringle 2001).}
\end{figure}
\begin{figure}
\plottwo{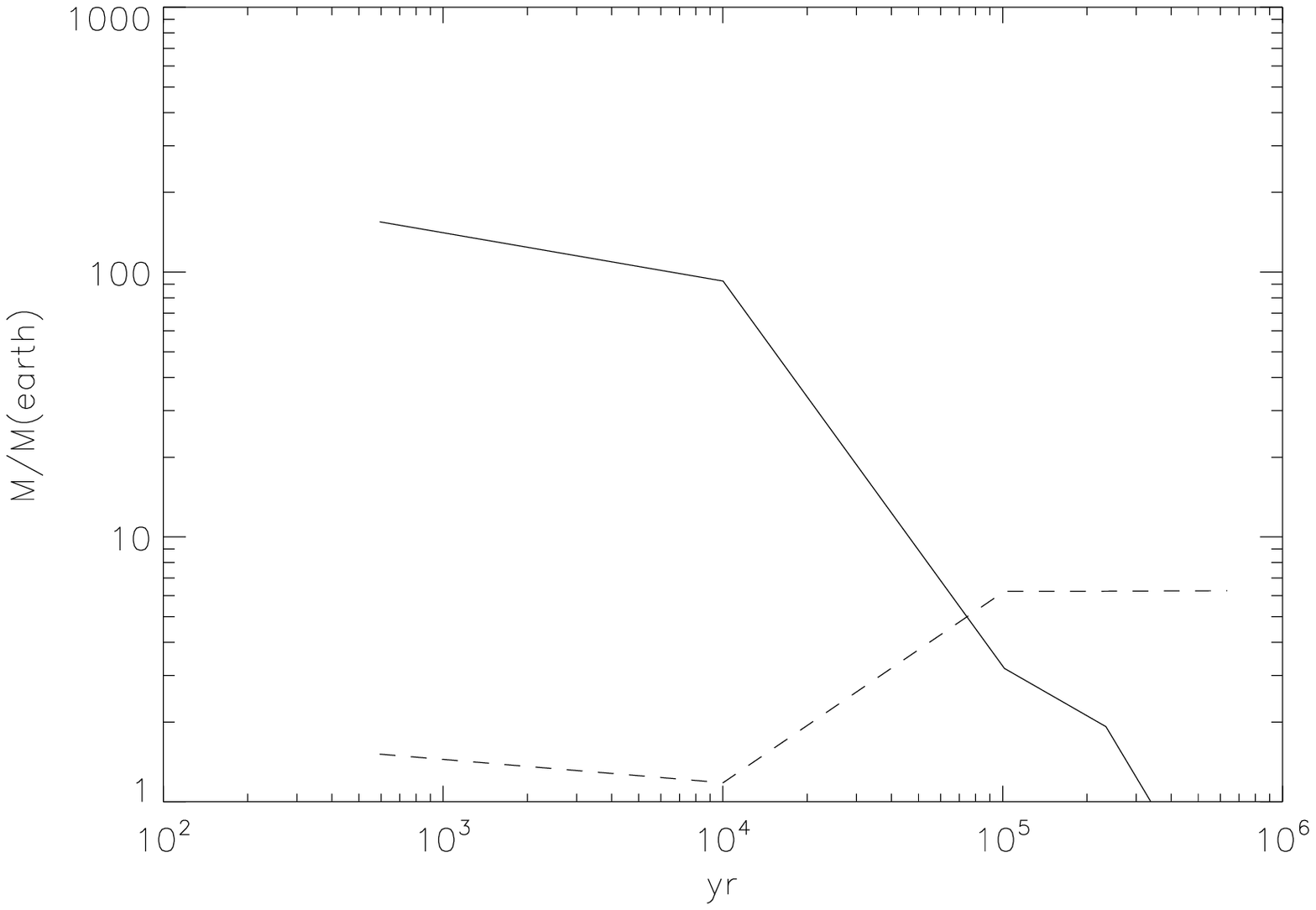}{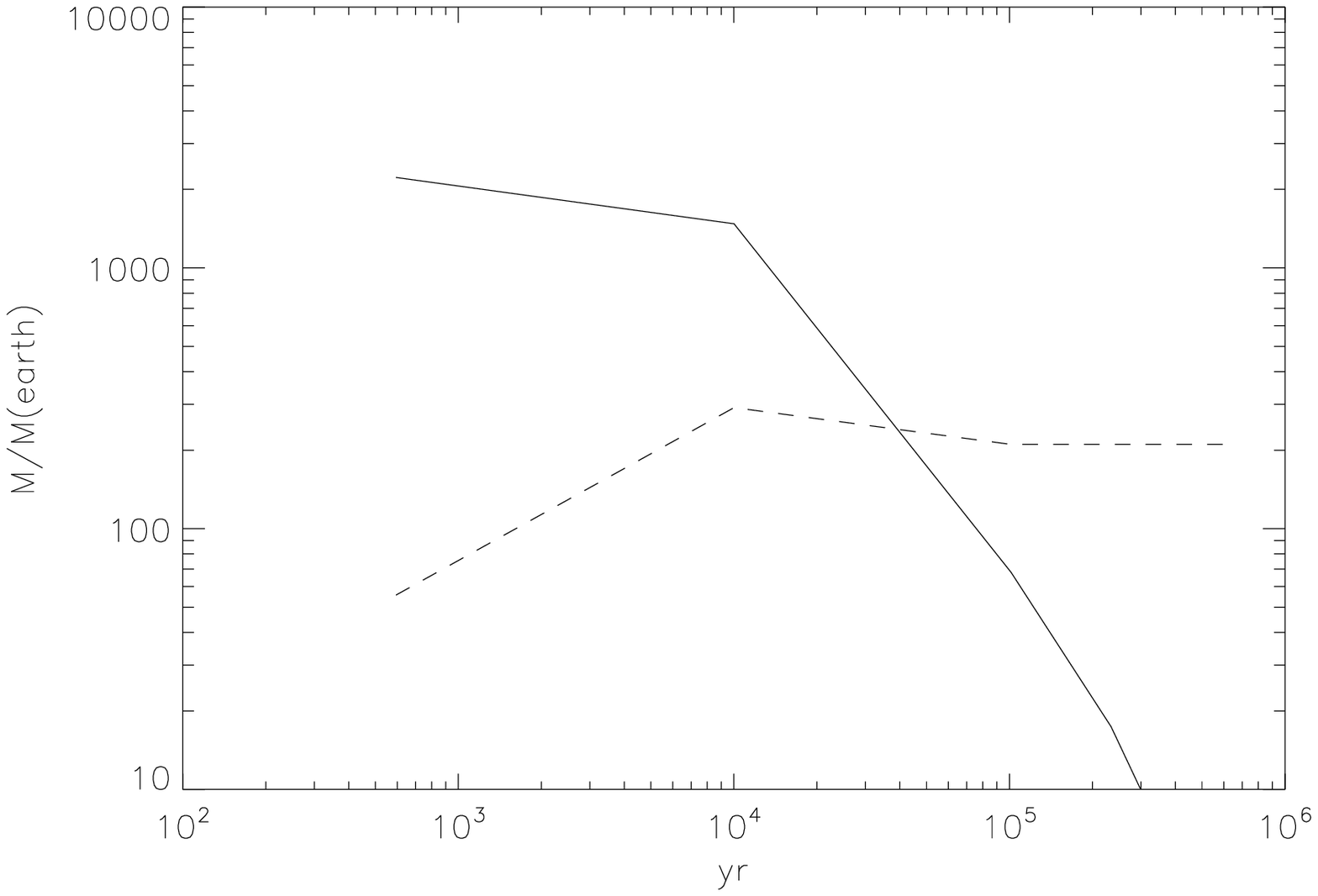}
\caption{The disk mass interior to 1 AU and exterior to 2 AU for the tidal-disruption disk model, assuming layered 
accretion an an initial angular momentum of $J\sim 10^{51}$ ergs$\cdot$s.  The final solid mass is reached very fast, 
within $2\times10^{5}$ and $10^{4}$ years for material $\le$ 1 and $\ge$ 2 AU, respectively.  The disk gas mass 
drops precipitously after 10$^{4}$ years.} 
\end{figure}
\begin{figure}
\plottwo{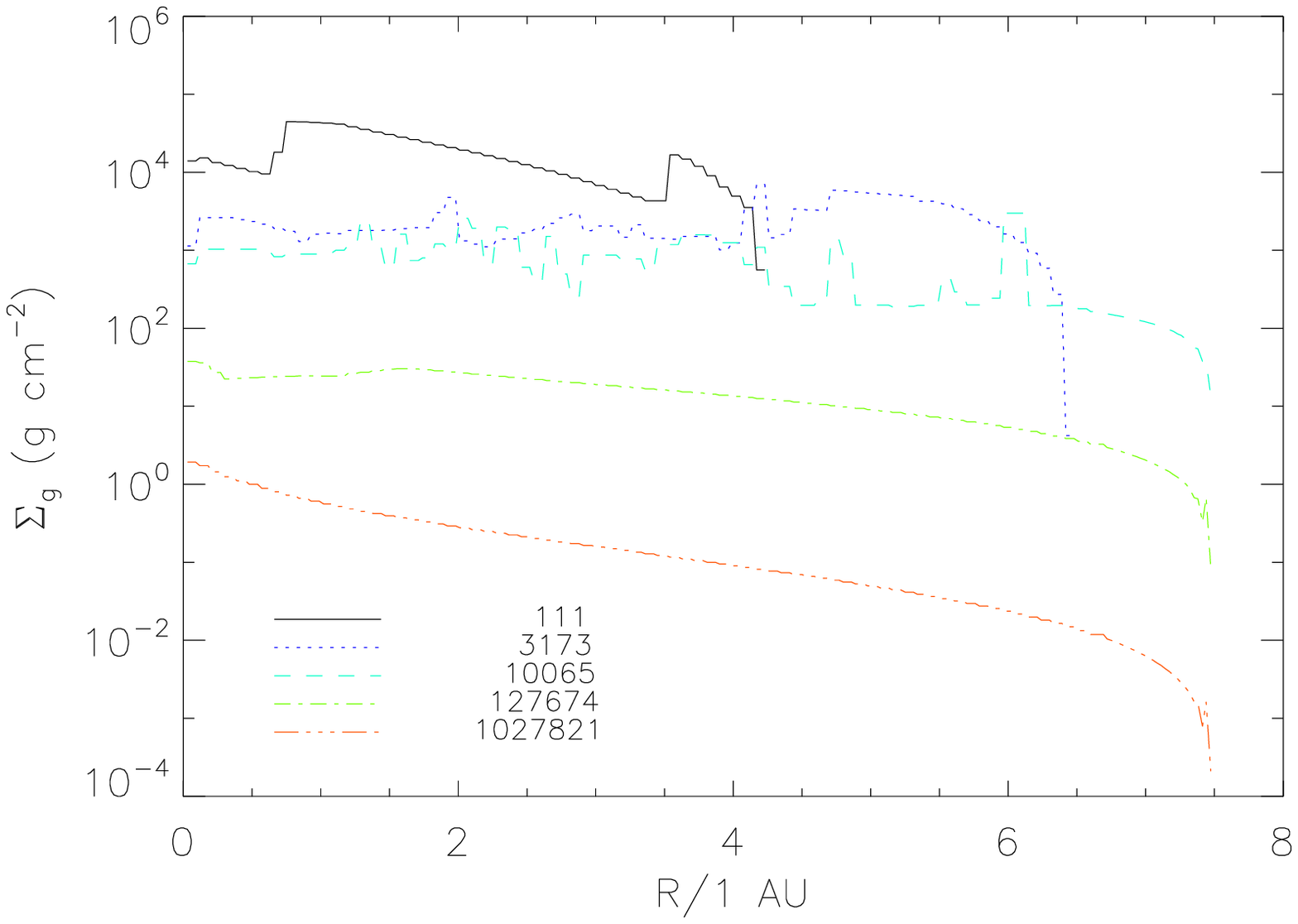}{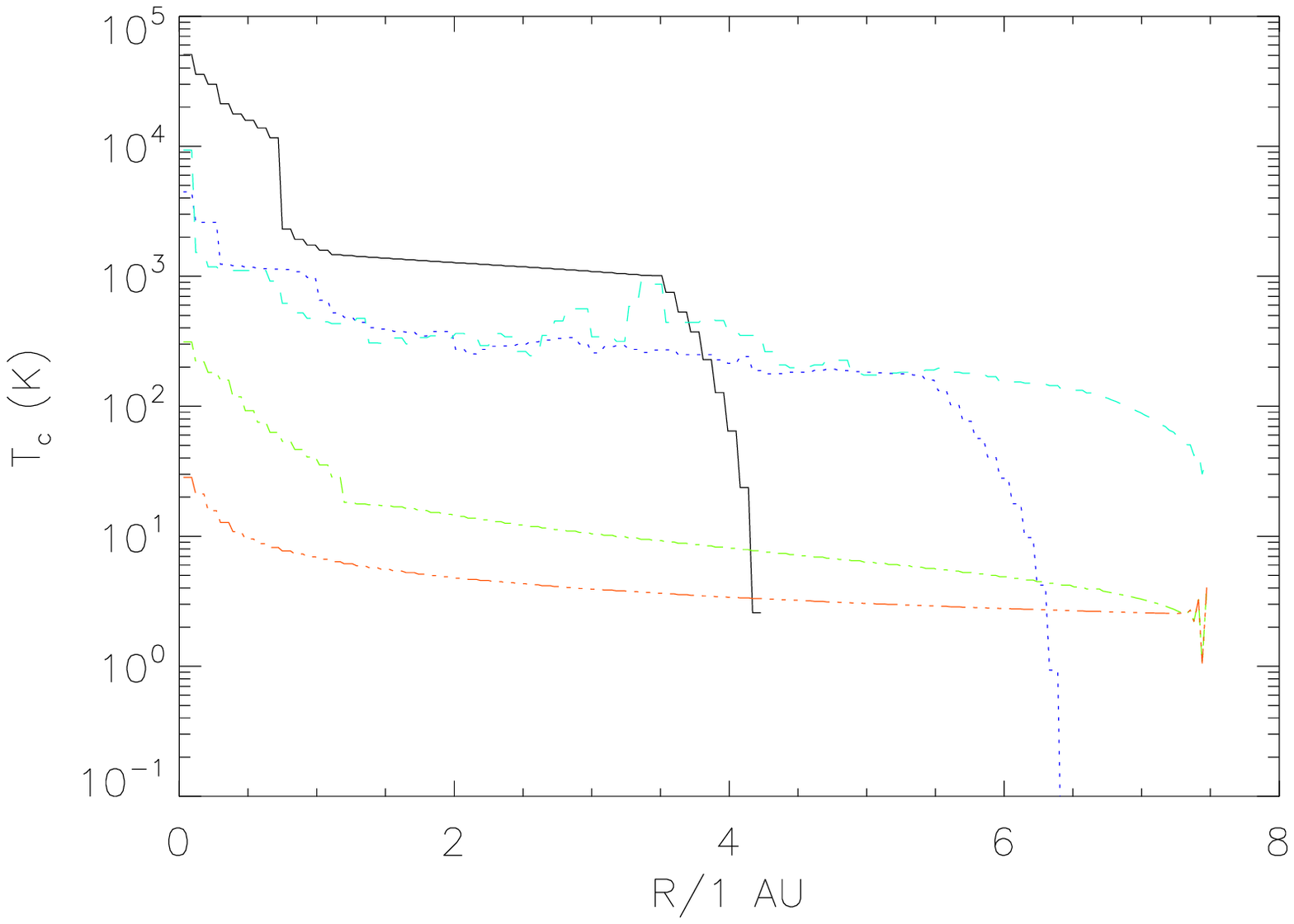}
\caption{Evolution of the gas surface density ($\Sigma_{g}$) and central temperature profiles for 
the tidal-disruption disk model, assuming layered accretion and an initial angular momentum of 
$J\sim 10^{52}$ ergs$\cdot$s.  The disk stays in a layered state from $10^{3}$ to $10^{5}$ years and evolves on 
$\sim$ $10^{5}$ year timescales. }
\end{figure}
\begin{figure}
\plottwo{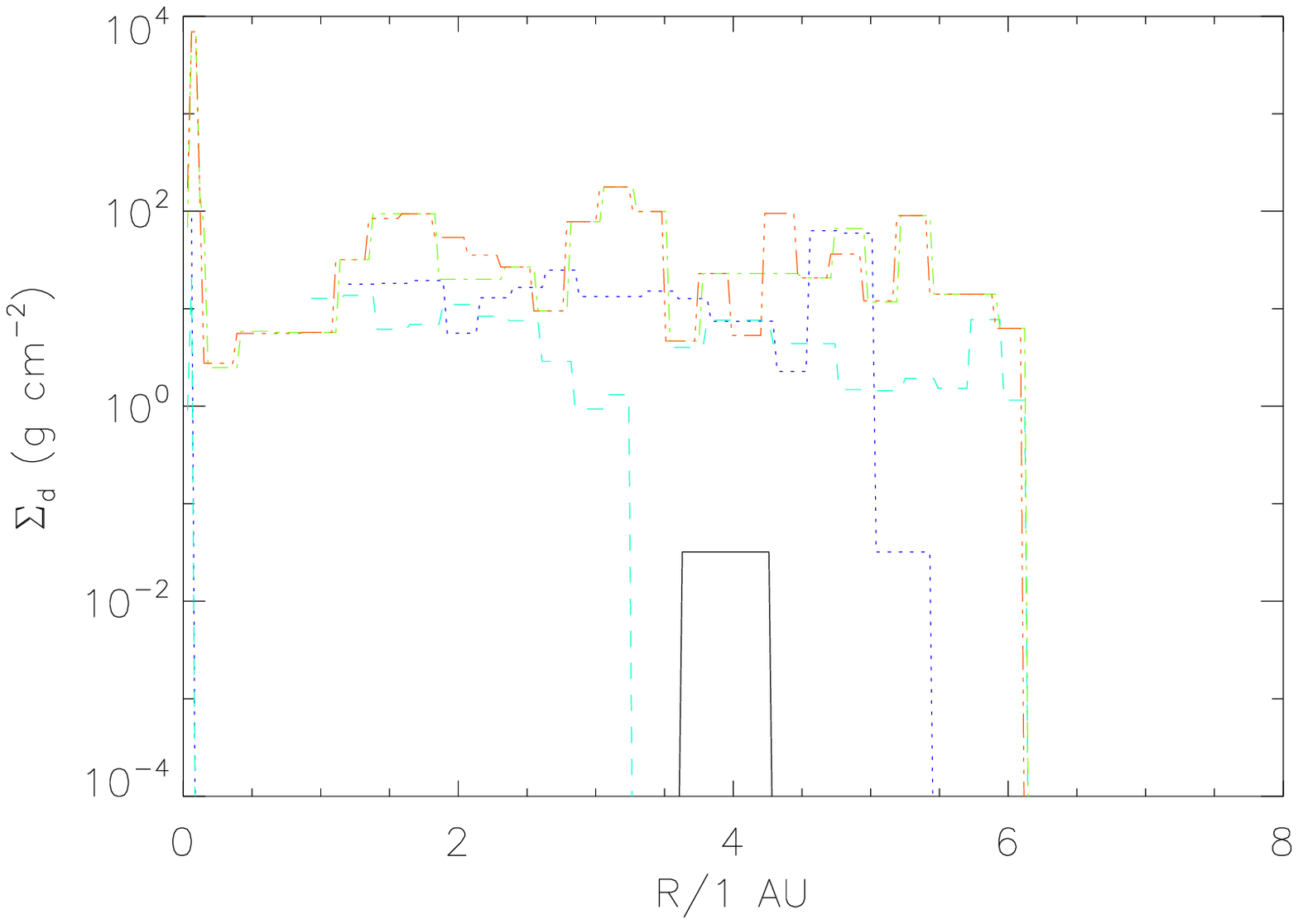}{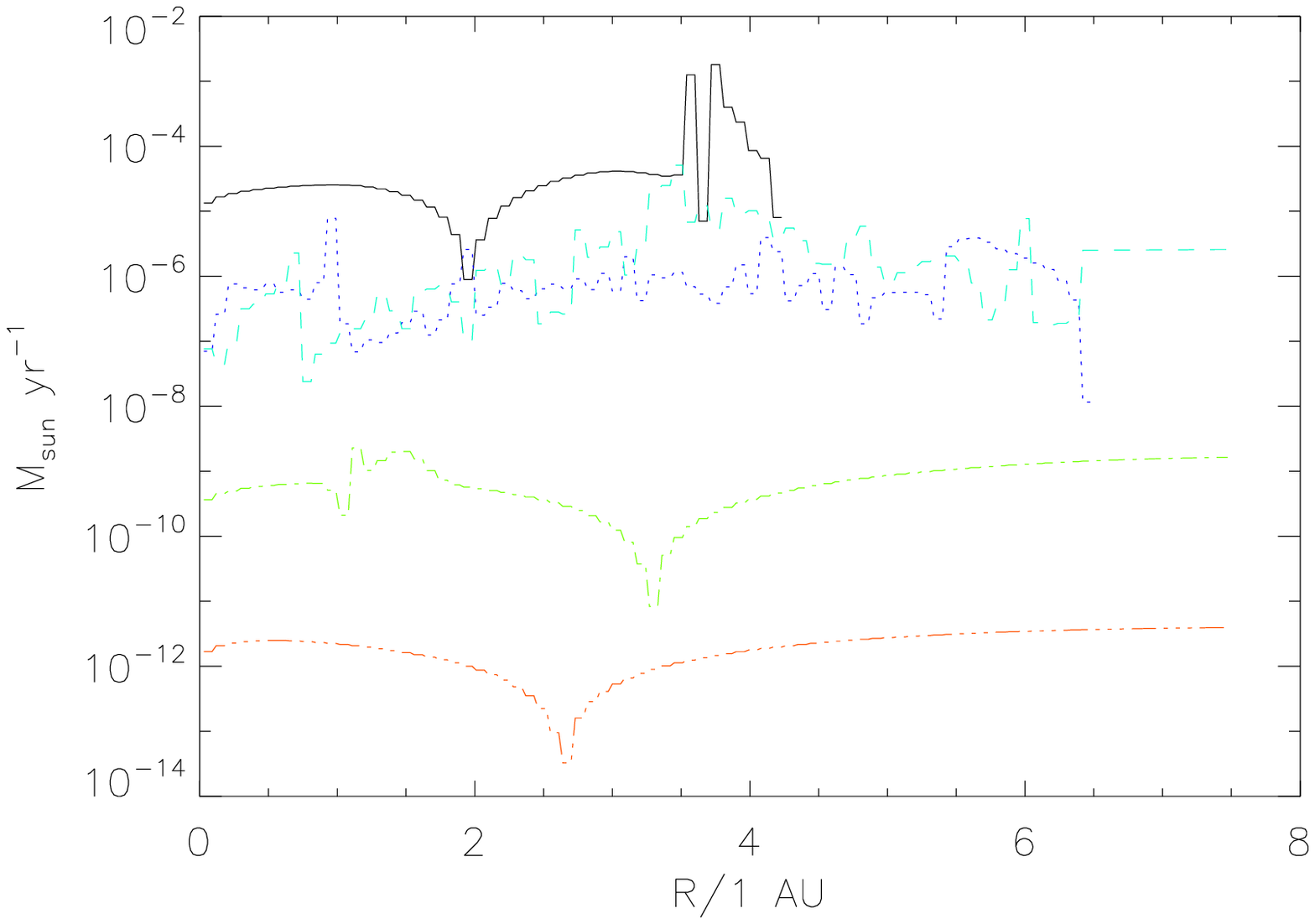}
\caption{Evolution of the solid surface density ($\Sigma_{d}$) and accretion rate profiles for
the tidal-disruption disk model, assuming layered accretion and initial angular momentum of
$J\sim 10^{52}$ ergs$\cdot$s.  The solid surface density ranges from $\sim$ 10-100 gcm$^{-2}$ from 
1-6 AU.  Less material is deposited beyond 6 AU and from 0.2-1 AU.
While the disk is in a layered state, the mass accretion rate also appears to be generally flat or increasing with 
distance as with the $J\sim 10^{51}$ case.}
\end{figure}
\begin{figure}
\plottwo{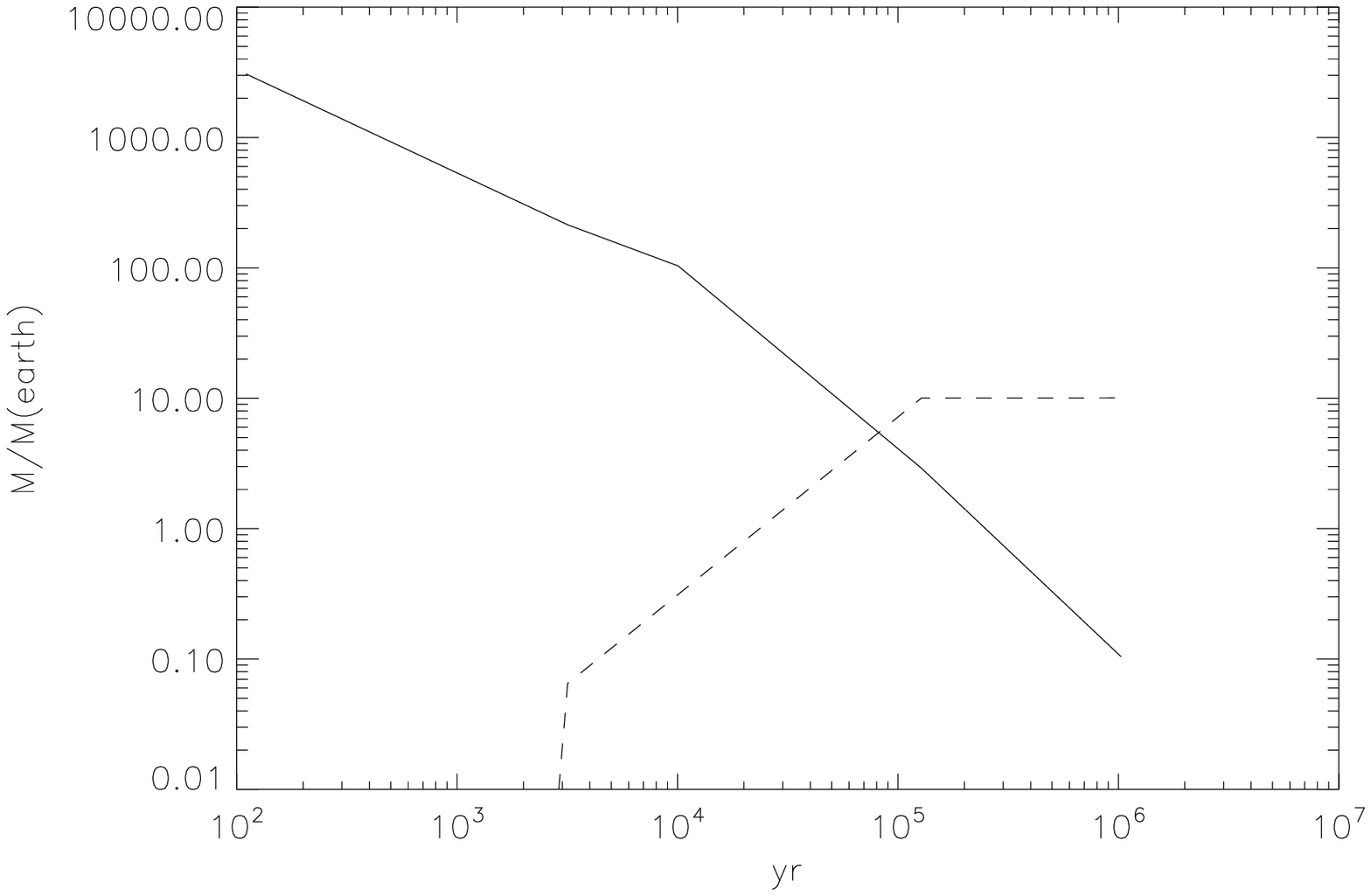}{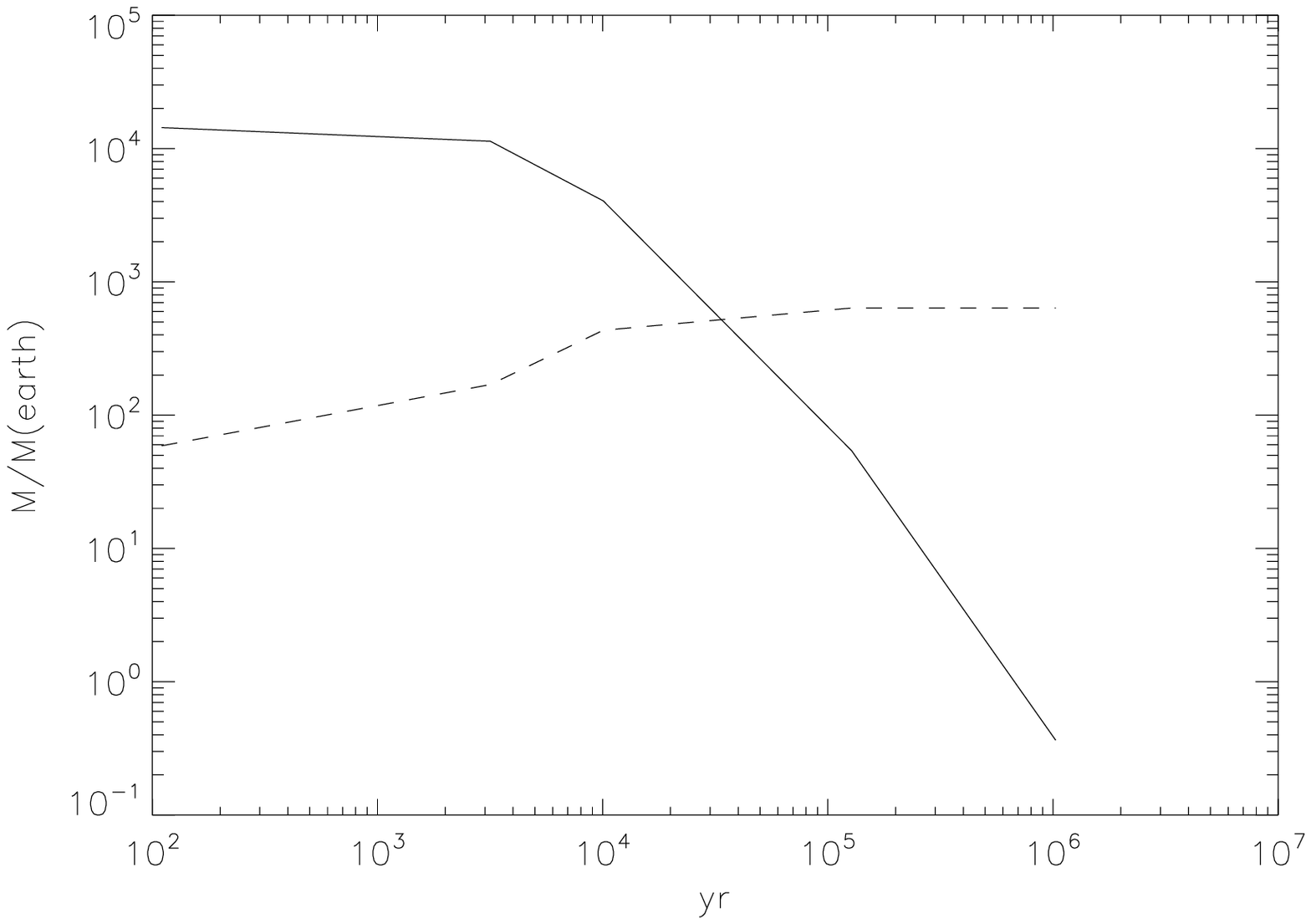}
\caption{The disk mass interior to 1 AU and exterior to 2 AU for the tidal-disruption disk model, assuming layered 
accretion an an initial angular momentum of $J\sim 10^{52}$ ergs$\cdot$s.  About 10 and 500 $M_{\odot}$ of solids 
emerge by $\sim$$10^{5}$ years interior to 1 AU and exterior to 2 AU, respectively.  However, 
by $\sim$ $10^{5}$ most of the gas has depleted where solids exist ($\lesssim$ 6 AU), making formation 
of gas giant planets highly unlikely.} 
\end{figure}

\end{document}